\documentclass[11pt]{article}
 \usepackage{amsmath}
 \usepackage{graphicx}
 \usepackage[merge,numbers,compress]{natbib}
 \usepackage[T1]{fontenc}
 \usepackage{booktabs}
 \usepackage{xcolor} 
 \usepackage{xspace}
 \usepackage{dcolumn}
 \usepackage{placeins}
 \usepackage[colorlinks=true,citecolor=blue!50!black,linkcolor=black]{hyperref}
 \usepackage{caption}
 \usepackage
 [subrefformat=parens,position=top,skip=-15pt,margin=15pt,justification=justified,singlelinecheck=false]
 {subcaption}
 \usepackage{multirow}
\usepackage{slashed}
\usepackage[bottom]{footmisc}
\usepackage{comment}
\usepackage{bigints}
\usepackage{braket}
\usepackage{amsfonts}
\usepackage{subcaption}
\usepackage{booktabs}
\usepackage[capitalise]{cleveref} 

\newcommand{\COM}{\sqrt{s_{max}}}

\newcommand{\rd}{\mathrm d}

\let\origfootnote\footnote
\renewcommand{\footnote}[1]{\kern.06em\origfootnote{#1}}
\newcommand{\punctfootnote}[1]{\kern-.06em\origfootnote{#1}}

\def\be{\begin{equation}}
\def\ee{\end{equation}}

\newcommand{\Pe}{\ensuremath{\text{e}}\xspace}

\newcommand{\Pt}{\ensuremath{\text{t}}\xspace}

\newcommand{\PW}{\ensuremath{\text{W}}\xspace}
\newcommand{\PZ}{\ensuremath{\text{Z}}\xspace}

\newcommand{\MW}{\ensuremath{M_\PW}\xspace}

\newcommand{\MeV}{\ensuremath{\,\text{MeV}}\xspace}
\newcommand{\GeV}{\ensuremath{\,\text{GeV}}\xspace}
\newcommand{\TeV}{\ensuremath{\,\text{TeV}}\xspace}

\newcommand{\abs}[1]{\left|#1\right|}

\newcommand{\newc}{\newcommand}
\newc{\bi}{\begin{itemize}}
\newc{\ei}{\end{itemize}}
\newc{\benu}{\begin{enumerate}}
\newc{\eenu}{\end{enumerate}}
\newc{\bc}{\begin{center}}
\newc{\ec}{\end{center}}
\newc{\bfig}{\begin{figure}}
\newc{\efig}{\end{figure}}
\newc{\qbar}{\bar{q}}
\newc{\go}{\tilde{g}}
\newc{\PB}{\textsc{Powheg-Box}}

\newcolumntype{.}{D{.}{.}{-1}}
\newcolumntype{d}[1]{D{.}{.}{#1}}

\colorlet{tableoverheadcolor}{gray!37.5}
\colorlet{tableheadcolor}{gray!25}
\colorlet{tablerowcolor}{gray!12.5}

\newlength{\width}
\newlength{\height}

\def\draftdate{\relax}
\def\mda{\relax}
\def\mua{\relax}
\def\mla{\relax}
\def\draft{
\def\thtystars{******************************}
\def\sixtystars{\thtystars\thtystars}
\typeout{}
\typeout{\sixtystars**}
\typeout{* Draft mode!
         For final version remove \protect\draft\space in source file *}
\typeout{\sixtystars**}
\typeout{}
\def\draftdate{\today}
\def\mua{\marginpar[\boldmath\hfil$\uparrow$]%
                   {\boldmath$\uparrow$\hfil}\color{black}%
                    \typeout{marginpar: $\uparrow$}\ignorespaces}
\def\mda{\color{red}\marginpar[\boldmath\hfil$\downarrow$]%
                   {\boldmath$\downarrow$\hfil}%
                    \typeout{marginpar: $\downarrow$}\ignorespaces}
\def\mla{\marginpar[\boldmath\hfil$\rightarrow$]%
                   {\boldmath$\leftarrow $\hfil}%
                    \typeout{marginpar: $\leftrightarrow$}\ignorespaces}
\def\Mua{\marginpar[\boldmath\hfil$\Uparrow$]%
                   {\boldmath$\Uparrow$\hfil}\color{black}%
                    \typeout{marginpar: $\uparrow$}\ignorespaces}
\def\Mda{\color{red}\marginpar[\boldmath\hfil$\Downarrow$]%
                   {\boldmath$\Downarrow$\hfil}%
                    \typeout{marginpar: $\downarrow$}\ignorespaces}
\def\Mla{\marginpar[\boldmath\hfil\textcolor{red}{$\Rightarrow$}]%
                   {\boldmath\textcolor{red}{$\Leftarrow $}\hfil}%
                    \typeout{marginpar: $\leftrightarrow$}\ignorespaces}
\overfullrule 5pt
\oddsidemargin 15mm
\marginparwidth 29mm
}

\newcolumntype{C}{>{\centering\arraybackslash}p{0.105\textwidth}}

\begin{document}

\title{\hfill ~\\[-30mm]
\phantom{h} \hfill\mbox{\small FR-PHENO-2025-02}
\\[1cm]
\vspace{13mm}   \textbf{A general approach to quantum integration \\ of cross sections in high-energy physics}}

\date{}
\author{
Ifan Williams$^{1\,}$\footnote{Corresponding author.}\,  \footnote{E-mail:
  \texttt{ifan.williams@quantinuum.com}}\, and
Mathieu Pellen$^{2\,}$\footnote{E-mail:
  \texttt{mathieu.pellen@physik.uni-freiburg.de}}
\\[9mm]
{\small\it $^1$ Quantinuum,} \\
{\small\it Terrington House, 13–15 Hills Road, Cambridge CB2 1NL, United Kingdom}\\[3mm]
{\small\it $^2$ Universit\"at Freiburg, Physikalisches Institut,} \\
{\small\it Hermann-Herder-Str. 3, 79104 Freiburg, Germany}
}
\maketitle

\begin{abstract}
\noindent
We present universal \emph{building blocks} for the quantum integration of generic cross sections in high-energy physics.
We make use of Fourier quantum Monte Carlo integration (MCI) as implemented in {\sc Quantinuum}'s quantum MCI engine to provide an extendable methodology for generating efficient circuits that can implement generic cross-section calculations, providing a quadratic speed-up in root mean-squared error convergence with respect to classical MCI.
We focus on a concrete example of a $1\to 3$ decay process to illustrate our work.
\end{abstract}

\thispagestyle{empty}
\vfill
\newpage
\setcounter{page}{1}

\tableofcontents
 \newpage

\section{Introduction}
Research in high-energy physics (HEP) aims at exploring fundamental physics at tiny scales.
To probe these small scales, elementary particles are collided with very high energies in order to produce other particles.
The prime example of such an experiment is the Large Hadron Collider (LHC), where protons are currently collided at a center-of-mass (COM) energy of $\sqrt{s} = 13.6\TeV$.
In this way, properties of known particles can be studied and new particles, such as the Higgs boson \cite{atlas12, cms12}, can be discovered.

In order to extract useful information from experimental data, theoretical predictions are needed for comparison.
At the heart of theoretical predictions is the concept of a \emph{cross section}.
Cross sections relate to the probability of a certain scattering process occurring in some collider experiment.
This means that all theoretical predictions are obtained through the computation of cross sections, which consist of either analytical or numerical integration.
As explained later, the method of choice for integrating a cross section in HEP is numerical Monte Carlo (MC) integration (MCI).
The computational resources needed to provide theoretical predictions for collider experiments are enormous; for the LHC, they are of the order of billions of CPU hours per year~\cite{Buckley:2019wov,HSFPhysicsEventGeneratorWG:2020gxw,CERN-LHCC-2022-005}.

It is therefore particularly interesting to look at whether quantum processing units can provide an alternative to CPU or GPU technology for computing theoretical predictions in HEP.
In the last few years, several works have been devoted to exploring the potential of quantum computing for computing theoretical predictions in perturbative theory~\cite{Bauer:2019qxa,Bepari:2020xqi,Perez-Salinas:2020nem,Bepari:2021kwv,Bravo-Prieto:2021ehz,Ramirez-Uribe:2021ubp,Deliyannis:2022uyh,Gustafson:2022dsq,Clemente:2022nll,Chawdhry:2023jks,Bauer:2023ujy,Ramirez-Uribe:2024wua}.
The quantum analogue of classical MCI is known as \emph{Quantum Monte Carlo Integration} (QMCI) \cite{montanaroMC, Herbert:2021xgs}, and this quantum algorithm utilises the \emph{Quantum Amplitude Estimation}~\cite{Brassard:2000} (QAE) algorithm as a key subroutine, providing a quadratic speed-up in root mean-squared error (RMSE) convergence with respect to classical MCI. 

In Ref.~\cite{Agliardi:2022ghn}, Mathieu Pellen (M.P.) investigated the use of QMCI to integrate elementary scattering processes. In this reference, a variant of the original QAE algorithm was used, namely \emph{Iterative QAE}~\cite{Grinko:2019} (IQAE).
This seminal work focused on simplified versions of scattering cross sections and explored both one- and two-dimensional integrals.
Following this, Refs.~\cite{deLejarza:2024pgk,deLejarza:2024scm} explored the same idea but using \emph{Quantum Fourier Iterative Amplitude Estimation}~\cite{deLejarza:2023qxk} for the quantum integration, which relies on the \emph{Fourier QMCI}~\cite{Herbert:2021xgs} (FQMCI) method (described later in \cref{sec:quantumapproach}). In Ref.~\cite{deLejarza:2023qxk}, a small-scale (small number of qubits) parametrised quantum circuit (PQC) is trained to prepare a quantum state representing a specific cross-section integrand, and decompose this state into its Fourier series. Each term in the Fourier series is then integrated separately using QAE, in the same manner as in FQMCI. A potential drawback of this method is that it may not scale efficiently to systems with larger numbers of qubits (as would be required for multi-dimensional integrands to be calculated, for example), due to the trainability issues associated with variational methods in quantum machine learning \cite{mcclean18,cerezo21,sharma22, wang21, marrero20, arrasmith20}. Indeed, preparing arbitrary probability distributions on a quantum computer is thought to be computationally hard in the general case, and the topic remains an active field of research.

In the present work, we go significantly beyond what has been done to date by presenting a methodology for the computation of generic cross sections in HEP, applicable to arbitrary dimensionality.
In particular, the method of quantum integration makes use of the FQMCI method implemented in {\sc Quantinuum}'s \emph{QMCI engine}~\cite{Akhalwaya:2023hqe} (developed in part by Ifan Williams [I.W.]). While research using the QMCI engine has so far been focused on financial applications \cite{cui2024uik}, it is particularly interesting to consider other potential fields such as HEP, where the integrals studied possess different challenges to those in finance \emph{e.g.}\ high dimensionality, more complex integrands, and non-trivial integral limits.

More concretely, in this article we investigate the computation of one- and two-dimensional cross-section integrations for scattering processes using the QMCI engine when run on a noiseless state-vector simulator. We demonstrate how one can decompose the integrand for an arbitrary, multi-dimensional cross section into products of constituent \emph{building blocks}, which are single-variable terms. We describe methods for implementing these individual building blocks on a quantum computer, using the in-built functionality of the QMCI engine. We also discuss how kinematical constraints can be applied to the integrals using this in-built functionality. Our findings demonstrate that this important future application of quantum computing is likely to require fully fault-tolerant hardware before it becomes practically useful.

In \cref{sec:def}, we begin by defining the problem and explain why numerical MCI is the preferred method for classical cross-section integration.
After this, we introduce FQMCI, {\sc Quantinuum}'s QMCI engine, and discuss how we can use the engine to perform QMCI for generic cross-section calculations.
We then discuss in detail some potential state-preparation methods for generating relativistic Breit-Wigner distributions---one of the key building blocks discussed---in \cref{sec:statepreparation}.
Some example applications demonstrating the applicability of our methodology are then discussed in \cref{sec:applications}. Finally, \cref{sec:conclusion} contains a discussion of the work, an outlook, and concluding remarks.

\section{Definition of the problem and approaches}
\label{sec:def}
\subsection{Cross sections in HEP}
\label{sec:XsectionHEP}

The cross section, typically denoted by $\sigma$, is a central concept in HEP.
It relates to the probability of a given scattering process $a+b\to c+d+\cdots$ occurring, where the letters represent elementary particles.
This value can be compared to experimental measurements.
In a general, abstract way, a cross section can be expressed as
\begin{align}
\label{eq:XS}
\sigma = \frac{1}{F} \int \rd \Phi |\mathcal{M}|^2 .
\end{align}
It is worth mentioning that for cross-section calculations, and thus for the calculations in this article, the integration is always performed over the real domain.
The \emph{flux factor}, $F$, characterises the rate at which the initial-state particles ($a$ and $b$) collide, and is a real number.
It is defined as
\begin{align}
F = 4 E_a E_b v_{\rm rel},
\end{align}
where $E_a$ and $E_b$ denote the energy of the initial particles, respectively, while $v_{\rm rel}$ is the relative velocity between them.
For fixed energies in the initial state, as is the case in this article, $F$ is a real constant, and thus can simply be factored out.
The phase-space term $\rd \Phi$ encompasses four-momentum conservation ($p_a + p_b = p_c+p_d+\cdots$) as well as the real integration over the kinematic variables of the final-state particles ($c,d,\ldots$).
Finally, the \emph{matrix element} $\mathcal{M}$ encodes the transition probabilities between the states $a,b$ and $c,d,\cdots$.
The matrix element itself is constructed from complex numbers and spinors, but when it is complex-conjugate squared [as in \cref{eq:XS}] it is always a real number.
The underlying theory of interactions between the particles is encoded in the matrix element.
In particular, when the interaction between the initial and final state is mediated via \emph{virtual} particles (not necessarily identical to the external ones), so-called \emph{propagators} will be present in the matrix element.
Their general form, arising in $|\mathcal{M}|^2$, reads
\begin{align}
\label{eq:propagator}
    p(x) = \frac{1}{(x-M^2)^2 + M^2 \Gamma^2} ,
\end{align}
where $M$ and $\Gamma$ represent the mass and the decay width of the massive particle, respectively.
It is worth mentioning that the decay width of an unstable particle is a measure of its decay probability per unit time. 
It is the inverse of the particle's lifetime, and in that respect, an important parameter for characterising a particle's dynamics.
For massless particles, the propagator in \cref{eq:propagator} reduces to $p(x)=1/x^2$, by setting $M=0$.
In this case, the variable $x$ is a kinematic invariant which has units of energy/mass squared.
This means that the general form of the integral to be performed in a cross-section calculation [as in \cref{eq:XS}] reads
\begin{align}
\label{eq:general_XS}
    \int \prod^{N_I}_{i=1} \rd x_i \frac{\sum_{S_k\in I} \alpha_{k} \prod_{j\in S_k} x^{n_j}_j}{\prod^{N_P}_{p=1} (x_p-M_{op}^2)^2 + M_{op}^2 \Gamma_{op}^2} ,
\end{align}
where $N_I$ and $N_P$ are the number of integration variables and propagators, respectively, so that $N_I\ge N_P$.
In the present notation, the variables of integration $x_i$ can either be kinematic invariants or angles related to the final-state particles.
The number of integration variables scales as $3n-4$ for a $2\to n$ scattering process.
The number of propagator terms depends on the process considered, which is defined by the initial- and final-state particles.
The index set $I$ contains all combinations of all the $N_I$ indices, which are denoted by the subset of indices $S_k$.
In this way, all possible monomials can be represented.
The index, $n_j$, represents the power of each term, and is a positive integer.
The labels $op$ denote all possible internal massive particles \emph{i.e.}\ all particles that are neither in the initial nor final state.
For example, for the process that will be studied later (see \cref{fig:Feynman}), 
the W boson is an internal particle.
Finally, $\alpha_{j\ldots k}$ are real constants to ensure full generality. There are as many constants as there are monomials.
The challenge is thus to integrate a multi-dimensional integral that has a highly peaked structure due to the propagators.\footnote{The propagators describe intermediate particles that can become resonant when the invariant mass of the four momentum is close to the mass of the particle. They are therefore sometimes referred to as \emph{resonances}.}
Note that the form of \cref{eq:general_XS} can be made even more general by allowing sums/differences of random variables instead of simply $x_l$ in the denominator. However, given that this case only arises at very high multiplicities (at least $1 \to 5$ or $2 \to 5$), we defer the treatment of this even more general case for future work.

To be more concrete, a $2\to 8$ process, such as the production of four leptons in association with four bottom quarks (which is dominated by the production of two top quarks in association with two bottom quarks)~\cite{Denner:2020orv}, will have $20$ integration variables, and $\mathcal{O}(1000)$ different propagators. 
Without going into the details of further complications related to considering higher orders in perturbation theory, this example shows the level of complexity for state-of-the-art theoretical computations in HEP.\footnote{The interested reader can look at Ref.~\cite{Huss:2022ful} and references therein to get an overview of the current frontier in theoretical predictions for HEP.}
It also explains why the computational resources needed for the LHC's physics programme is so large.\footnote{To be exact, the computational time quoted in the introduction is for the generation of so-called \emph{events}, representing the underlying distribution to be integrated.
These events are thus closely related to the integration of the cross section.}

A further complication when computing cross sections is the appearance of event selections in experimental analyses.
Typically, the detectors of an experiment cannot physically cover the entire phase space. Therefore, the integration in \cref{eq:XS} has to be restricted to a smaller domain of integration, corresponding to that physically accessible by the experiment.
This is done by setting minimal and maximal values (setting the cut values) to observable quantities, such as the rapidity or the transverse momentum of the final particles.\footnote{The rapidity of a particle is defined as $y=\frac12 \log\left( \frac{E+p_{\rm L}}{E-p_{\rm L}} \right)$, where $E$ is the energy of the particle and $p_{\rm L}$ its momentum along the beam axis, typically taken as the $z$ axis by convention. With this  convention, the transverse momentum is defined as $p_{\rm T} = \sqrt{p_x^2+p^2_y}$.}
Events with particles that do not pass these requirements are not detected.
From the theoretical side, these observable quantities depend on the final-state momenta, which themselves depend on the integration variables.
The cross section therefore becomes
\begin{align}
\label{eq:xsection_cuts}
    \sigma = \frac{1}{F} \int \rd \Phi \, |\mathcal{M}|^2 \, \Theta\left(C[\Phi]-C[\Phi_c]\right) ,
\end{align}
where the \emph{cut function}, $C$, represents these restricted observables. Formally, the function $C$ depends on the variables of integration, as it depends on $\Phi$ (the final-state particles), which is itself built from the variables of integration. However an analytical form for $C$ as a function of the variables of integration may not necessarily exist; we will not investigate this aspect in detail in the rest of the article, however we will briefly discuss it. 

We now move on to discuss classical approaches for calculating such integrals.

\subsection{Classical approach}

The most common approach to compute integrals such as the ones presented in the previous section is by reverting to numerical MCI techniques.
There are several reasons for this:
first, as eluded to above, closed forms for the cut functions do not always exist. This means that numerical techniques must be used.
Second, while analytical integration is possible for simple scattering processes (\emph{e.g.}\ $2\to 2$ or $2\to 3$), with increasing multiplicity the calculations rapidly become intractable.
Third, reverting to analytical calculations make things more difficult to automate, as there are typically many processes that are relevant for a given experiment, \emph{e.g.}\ $\mathcal{O}(100)$ different processes for the LHC.
The LHC in particular also has the drawback that it is a proton-proton collider, meaning that calculations rely on knowledge of the \emph{parton distribution functions} for the proton, functions that provide the probability for a parton (an elementary particle such as a quark or gluon) to be emitted.
These functions are not known from first principles, and must be extracted from data.
As they are typically defined in terms of numerical grids, this makes them more naturally embeddable within a numerical framework.
Finally, while one is primarily interested in computing the cross section, this quantity is also measured as a function of other experimental observables such as angles between particles or observables related to energies. When using MCI, computing the cross section as a function of other observables comes at no extra cost,
as at each evaluation, the weight used for the cross-section evaluation can also be binned in a histogram according to the value of the observable of interest.
In this case, while the binning of the histograms is a computationally cheap operation,
for analytical calculations this would require additional, more complicated calculations.

The RMSE of classical MCI scales as $\mathcal{O}(1/\sqrt{\mathcal{S}})$, where $\mathcal{S}$ is the number of samples.
Improvements on basic MC techniques aim therefore to reduce the overall variance. This is done either by modifying the original integrand so that the new function has a smaller variance, or by sampling the integration such that the distribution of sampled points is no longer randomly uniform across the entire domain of integration~\cite{Byckling:1971vca}.

The first type of technique is known as \emph{importance sampling}.
For HEP, propagators are mapped to functions with lower variance.\footnote{See Ref.~\cite{Byckling:1971vca} for a pedagogical explanation in the case of a single propagator.}
Given that the propagator structure is known a priori (thanks to knowledge of the Feynman diagrams that contribute to the process), it is possible to setup an integration routine such that each of the resonant structures are singled out.
This leads to the so-called \emph{multi-channelling} integration technique.

The second type of approach, is typically referred to as the \emph{adaptive} MC technique~\cite{Lepage:1977sw}.
The idea is that, through an iterative procedure, points are sampled according to the distribution to be integrated. This means that the peak structure is probed more efficiently.

For state-of-the-art integration in HEP, in reality combinations of both approaches are utilised.
Nonetheless, the scaling of the overall RMSE is still $\mathcal{O}(1/\sqrt{\mathcal{S}})$, motivating the exploration of quantum algorithms for sampling, which are known to have better scaling.\footnote{It is fair to say that \emph{Markov-chain MC} (MCMC) and \emph{quasi-MC} approaches have so far received little attention as methods for cross-section integration in HEP.
While the former seems promising~\cite{Kharraziha:1999iw,Kroeninger:2014bwa,Yallup:2022yxe,LaCagnina:2024wcc}, the latter is known to be more appropriate for calculations at higher orders of perturbation theory~\cite{Li:2015foa,deDoncker:2018nqe,Borowka:2018goh}.
The reason is that quasi-MC methods are not appropriate for integrands with highly peaked structures, and where there are asymmetric restrictions on the phase space---as is the case for cross-section calculations.}

In order to fix notation for later use and clarity, we here define the problem of MCI in a more rigorous way.
MCI calculates the expectation of a function, $g(x)$, of a continuous random variable, $X$, with probability density function (PDF), $f_{X}(x)$, defined as
\begin{equation}\label{eq:functionappliedintergral}
    \mathbb{E}\left[g(X)\right] = \int g(x)f_{X}(x)\,\rd x.
\end{equation}
We will hereafter refer to $g(.)$ as the \emph{function applied}. We can straightforwardly extend this to functions, $g(x,y)$, of two continuous random variables, $X$ and $Y$, with joint PDF, $f_{XY}(x,y)$,
\begin{equation}\label{eq:functionappliedintergral2D}
\mathbb{E}\left[g(X,Y)\right] = \iint g(x,y)f_{XY}(x,y)\,dxdy.
\end{equation} Indeed, this can extended further to an arbitrary number of dimensions, and with the integration performed over either all or subsets of the given variables. However, for the purposes of this article it will only be necessary to focus on expectations of separable functions of at most both $X$ and $Y$ \emph{i.e.}\ products of functions, $g(x,y)=h(x)l(y)$, with expectation
\begin{equation}\label{Eq:FunctionAppliedIntergral2Dsplit}
\mathbb{E}\left[h(X)l(Y))\right] = \iint h(x)l(y)f_{XY}(x,y)\,dxdy.
\end{equation}
We now move on to discuss the equivalent approach in quantum computing.
\subsection{Quantum approach}\label{sec:quantumapproach}

The quantum analogue of classical MCI, QMCI, is a quantum algorithm that returns a numerical estimate for the value of some (possibly multidimensional) integral, in the same manner as for classical MCI. 

QMCI provides a quadratic speed up in the convergence of the RMSE of the estimate as a function of the number of samples, as compared to classical MCI.
Therefore, the RMSE scales as $\mathcal{O}(1/\mathcal{S})$, compared to $\mathcal{O}(1/\sqrt{\mathcal{S}})$ classically. The source of the quantum advantage in QMCI arises from the use of QAE~\cite{Brassard:2000}---a generalisation of \emph{Grover's Search} algorithm \cite{grover1996rk}---which is the key subroutine of QMCI.

The QMCI algorithm comprises three main steps, as described in Ref.~\cite{Akhalwaya:2023hqe}, where for demonstration purposes we only consider a single random variable, $X$, (however this is straightforwardly extended to multiple variables).
\begin{enumerate}
\item A quantum state, $\ket{p}$, encoding a multivariate probability distribution, $f(.)$, in the values of its complex amplitudes is prepared by applying a quantum circuit, $P$, such that 
\begin{align}
\ket{p} = P \ket{0}_n = \sum_{x} \sqrt{f_X(x)} \ket{x}. 
\end{align}
\item An observable function, $g(.)$, is applied coherently to the state based on a quantum arithmetic circuit, $R$, with the expectation value of interest, $\mathbb{E}\left[g(X)\right] = \sum_{x} g(x)  f_X(x)$, then encoded in the amplitude of an additional ancilla qubit as
\begin{align}
R \ket{p} \! \ket{0} = \sum_{x } \sqrt{f_X(x)} \ket{x}  \left( \sqrt{1-g(x)} \ket{0} + \sqrt{g(x)} \ket{1} \right).
\end{align}
\item The expectation value of interest---which is equal to the probability of measuring the 1 state---is estimated using QAE.
\end{enumerate}

At this stage, it is also worth briefly discussing the fact that the first step of QMCI necessarily entails some state-preparation protocol, which has an associated computational cost. A natural question to ask, then, is how the state-preparation overhead affects the overall quadratic advantage of QMCI, and under what assumptions this advantage is fully retained. Herbert \cite{herbert21} discusses this in detail, and notes that the computational advantage is retained if the operational cost of preparing the quantum state encoding the probability distribution is on par with that of generating a classical sample---an efficiency achieved by the $Q$-marginal construction, which allows such quantum states to be prepared directly from reversible classical sampling circuits using only a single layer of Hadamard gates.

\subsubsection{Fourier Quantum Monte Carlo Integration}
\label{sec:fqmci}

In Ref.~\cite{Herbert:2021xgs}, Herbert proposed the FQMCI methodology as an efficient (low depth) means of performing QMCI that retains the full quadratic advantage without requiring costly quantum arithmetic. The idea stems from the fact that $\mathbb{E}[\text{sin}^{2}(X)]$ is a straightforward quantity to estimate on a quantum computer (merely requiring a simple circuit corresponding to a bank of $R_y$ rotation gates, rather than any complicated quantum arithmetic). In FQMCI the function applied---under the assumption that it obeys certain smoothness conditions (continuous in value and first derivative, and second and third
derivatives piecewise continuous and bounded)---is extended as a periodic, piecewise function, and this is then decomposed as a Fourier series. Then, because each term in the series is a cosine or sine, their expectations can be easily estimated individually, and these then recombined. 

This methodology is also easily extended to the bivariate case, where there are expectations of products of functions applied of the form $\mathbb{E}[h(X)l(Y)]$. In this case the individual Fourier series for $h$ and $l$ are multiplied together. One then ends up with univariate sine and cosine terms, alongside bivariate products of sines and cosines, which in the latter case can be reduced to sums of single bivariate sines or cosines using trigonometric identities. Thus the expectation can again be estimated by calculating expectations of sines and cosines, and then recombining. However, we note that this differs from the univariate case as multiple expectations are required to be calculated, and thus the resources required to perform the calculation are greater. In addition, the circuits are larger, because the expectations are for bivariate trigonometric functions (e.g, $\mathbb{E}[\text{sin}^{2}(X-Y)]$), requiring an additional bank of $R_y$ rotation gates to represent the $Y$ variable.

It is important to note for what follows that this technique can also be extended to an arbitrary multivariate case \emph{i.e.}\ to expectations of products of functions applied of the form $\mathbb{E}[h(X)l(Y)m(Z)...]$. However, for products of more than two this technique is unlikely to be particularly efficient due to the polynomial increase in the numbers of terms required to compute, and the additional increase in size of the circuits involved.

\subsubsection{{\sc Quantinuum}'s QMCI engine}
\label{sec:qmci_engine}

{\sc Quantinuum}'s QMCI engine \cite{Akhalwaya:2023hqe} is the world's first fully integrated platform for performing QMCI, with each of the steps discussed previously implemented within the engine based on state-of-the-art methods. There are a number of optimised, in-built features that will become important for the later discussion of how to implement generic cross-section calculations.

First, the engine can take as input any circuit, $P$, provided by the user, and then this circuit is straightforwardly passed along the rest of the QMCI pipeline. It also contains an in-built \emph{state-preparation library} (the \emph{distribution loader}) containing circuits that represent some common distributions, such as the uniform and Gaussian distributions (see Section 7 of Ref.~\cite{Akhalwaya:2023hqe} for a detailed description of the distribution loader). Second, based on the FQMCI method, the engine provides protocols to implement efficient Fourier-decomposition circuits, $R$, for a variety of functions applied, particularly corresponding to moments or products of moments of random variables, such as for $g(x) = x^2, g(x,y) = xy$ etc.\ (see Section 8 of Ref.~\cite{Akhalwaya:2023hqe} for a detailed description of the types of quantities that can be estimated using the QMCI engine). Third, the engine contains a number of in-built algorithms for performing QAE, in general optimised for performance\footnote{This optimisation generally involves setting the number of shots such that the RMSE of the estimator is minimised.} (see Section 9 of Ref~\cite{Akhalwaya:2023hqe} for a detailed description of the in-built QAE algorithms). Fourth, the engine is able to implement limits on integral calculations, based on the use of \emph{thresholding operations}, implemented via the \emph{enhanced $P$-builder} functionality. Thresholding operations correspond to binary operations based on some threshold value, $V_{th}$, that enable the estimation of $\mathbb{E}\left[X\,\Theta(X \geq V_{th})\right]$, where the indicator function, $\Theta(z)$, with $z$ either a random variable or an expression, is defined as\footnote{The threshold is enforced by adding the one’s complement of the classical threshold value, ensuring that the most significant result qubit is set when the condition is met. A subsequent step resets the remaining result qubits to the zero state while preserving the most significant qubit. This makes the cleared qubits available as ancillas for future operations.}
\begin{equation}
    \Theta(z) = \begin{cases}
        1, & \text{if}~ z \ge 0 \\
        0, & \text{otherwise}.
    \end{cases} \label{eq:indicator_var}
\end{equation}
Fifth, the user is able to perform a QMCI calculation such that the RMSE of the final estimate is upper bounded by a chosen value (see Section 8 of Ref.~\cite{Akhalwaya:2023hqe} for a description of this). Finally, the engine is equipped with a \emph{resource mode} that can determine the quantum resources (in terms of qubit numbers and gate counts/depths) required to perform a given QMCI calculation to a desired RMSE. The resource calculator outputs either: with a view to running calculations in the noisy intermediate-scale quantum (NISQ) era---where applying two-qubit gates are the main bottleneck for hardware---the total number and depth of $CX$ gates across all circuits, alongside the same information for the largest circuit that is run, and the same information for the total number and depth of all gates; or, with a view to running calculations in the fault-tolerant era with full quantum error correction---where applying $T$ gates will be the main bottleneck for hardware---the total number and depth of $T$ gates across all circuits, alongside the same information for the largest circuit that is run. In both cases, the total number of qubits required for the largest circuit is also output (see Section 10 of Ref.~\cite{Akhalwaya:2023hqe} for a detailed description of exactly how the resource mode is implemented). Overall, these features will be key to later discussions in this article.

We note that how to split the integrand for the expectation value into the product of function applied, $g(x)$, and probability distribution, $f_{X}(x)$, is in general arbitrary. For QMCI applications in finance, which is so far the only application that has been previously studied using the QMCI engine \cite{cui2024uik}, there is a clear notion of calculating some expectation of the random variable for use in a financial calculation, such as the mean, $g(x)=x$, or the second moment, $g(x)=x^2$. However, for general applications (\emph{i.e.}\ integrands) there is freedom in how to perform this split, and we take advantage of this notion in the next section.

\subsection{Implementing a generic cross-section calculation}
\label{sec:generalform}

In the introduction, we have mentioned so called generic \emph{building blocks} for the computation of cross sections.
To illustrate this, we provide here two concrete examples.
From \cref{eq:XS,eq:general_XS}, a cross section with only a single integration variable reduces to
\begin{align}
    \sigma_1 \propto{} \int_0^s \frac{\rd x \, x^n}{\left(x-M_o^2\right)^2+M^2_o \Gamma_o^2} ,
\end{align}
where $s$ is the maximal value of $x$.
In order to cast the above equation into the more rigorous one of the MCI problem as in \cref{eq:functionappliedintergral}, noting that we are free to choose how to split the integrand for the expectation value into the product of function applied, $g(x)$, and probability distribution, $f_{X}(x)$, it follows that we can choose the function applied to be
\begin{align}
    g(x) = x^n,
\end{align}
(\emph{i.e.}\ the numerator). The probability distribution to sample from can then be associated to the propagator function (\emph{i.e.}\ the denominator), such that
\begin{align}\label{eq:bw_distribution}
    f_X(x) = \frac{1}{\left(x-M_o^2\right)^2+ M_o^2\Gamma_o^2} .
\end{align}
The reason we chose this particular factorisation is twofold. First, as mentioned previously, implementing moments or products of moments of random variables as the function applied is straightforward to implement based on the QMCI engine's existing functionality, and the RMSE scaling of the corresponding Fourier series' are likely favourable as compared to those of more complex functions. Second, because the probability distribution corresponds to a real distribution that has physical significance  (being used to model resonance behaviour in general), we can envision that future development of, for example, specific state-preparation methods tailored to this distribution\footnote{Note that in this article we only consider general state-preparation methods for preparing the BW distribution, and we defer research into more tailored methods for future work.} would not only be useful for this particular application, but also for future research directions in HEP (\emph{e.g.}\ for quantum-computing applications where resonance behaviours are required to be modelled).

We can identify the general form of \cref{eq:bw_distribution} as an (unnormalised) relativistic BW distribution~\cite{Breit:1936zzb} (hereafter simply denoted by BW distribution). As mentioned previously, state preparation of arbitrary probability distributions is a difficult problem. However, in this case we need only prepare one specific distribution, and as will be discussed in the following section, it will suffice to prepare this distribution a single time (for a given COM energy) for each of the (small number of) relevant massive particles in the Standard Model, \emph{i.e.,}\ the $\PW$ and $\PZ$ bosons, the top ($\Pt$) quark, and the Higgs boson.
In the present example, the two building blocks are the function, $g$, which is a monomial, and the BW distribution, $f_X$.

While not the focus of this article, it is also worth briefly discussing the case of massless particles where, as discussed previously, the propagator term then reduces to 
\begin{align}
    p(x) = \frac{1}{x^2} .
\end{align}
Such terms would then be absorbed in the monomials in the numerator, possibly with negative exponents. For this case of negative exponents, because the functions applied for FQMCI must obey certain smoothness conditions, as detailed previously in \cref{sec:fqmci}, then in this case one would not be able to exploit FQMCI. Instead, an approach similar to those of prior works would be required, where the entire integrand is prepared as the probability distribution and then QAE run directly on this state (equivalent in our language to the `function applied' being constant). 

In the same way, a general expression for a two-dimensional integration reads
\begin{align}
    \sigma_2 \propto{} \iint \frac{\rd x \rd y \, x^n y^m}{\left[\left(x-M_{o1}^2\right)^2+M^2_{o1} \Gamma_{o1}^2\right]\left[\left(y-M_{o2}^2\right)^2+M^2_{o2} \Gamma_{o2}^2\right]} ,
\end{align}
where the term $x^n y^m$ means that the numerator can be made of any monomials which are a product of $x$ and $y$ raised to various powers.
It implies that in order to match the form of \cref{Eq:FunctionAppliedIntergral2Dsplit}, the separable functions applied should be
\begin{align}
h(x)=x^m , \quad\quad l(y)=y^m.
\end{align}
Then, the probability distribution becomes
\begin{align}
    f_{XY} (x,y) = \frac{1}{\left[\left(x-M_{o1}^2\right)^2+M^2_{o1} \Gamma_{o1}^2\right]\left[\left(y-M_{o2}^2\right)^2+M^2_{o2} \Gamma_{o2}^2\right]}.
\end{align}
In the case that we consider, it should be noted that the two BW distributions factorise in the sense that each propagator is a function of a single random variable.
This property is particularly useful as it means that each propagator can be associated to one particular random variable, allowing the propagator product to be cast as a  simple product of orthogonal propagator terms. This is key, as it means that our approach is easily extendable to multiple dimensions, as would be required in the general case.

The strategy is thus to consider as the function applied the monomials in the numerator, while the propagators terms are treated as the probability distribution to be sampled from. 
Analogously, the basic building blocks are simply monomials, and BW distributions.
As highlighted here, this approach can be extended to arbitrary dimensions. 

The QMCI engine provides efficient Fourier decomposition circuits for moments or products of moment of random variables, as discussed previously, and allows the user to provide the probability distribution to be sampled from as an input. In addition, when constraints on the phase space are required to be imposed on the integration, as given by the cut function, $C$, in \cref{eq:xsection_cuts}, then for simple constraints such as sums, maximum/minimum etc.\ the engine's built-in thresholding operations can be used to achieve this. Thus---providing there are efficient ways to prepare BW distributions as quantum states, something discussed in the next section---in principle the QMCI engine can be used to perform quantum integration of a generic cross section in HEP.

It is worth noting that the cut function, $C$, is a function of the integration variables, and as discussed previously, possibly non-analytical. This means that in order to implement the restriction of phase space, the value $C(\Phi)$ should be numerically evaluated at each phase-space point. Such an approach has been followed in \cref{sec:applications} to set the boundaries of integration of a given variable that depends on another variable of integration; however, in this case the expression was analytical, and the quantum arithmetic implemented via the QMCI engine's enhanced $P$-builder. For the case of some non-analytical form, one possibility is to implement the required quantum arithmetic by constructing the equivalent classical circuit, and then making it a reversible circuit. However, we note that this approach is unlikely to be efficient, and indeed, in the future there may be better ways of constructing such circuits. We leave this aspect for future work.


\section{State preparation of relativistic BW distributions}
\label{sec:statepreparation}

We see that to be able to perform generic cross-section calculations using the QMCI engine, we require a methodology for preparing on a quantum register BW distributions for the massive propagators $\PW$, $\PZ$ and $\Pt$,\footnote{In practice, completeness would also require a propagator for the Higgs boson. However, due to the narrow width of the resonance, generating this peak with sufficient resolution is highly challenging, given current qubit counts and methods.
At high-energy colliders, all leptons are typically considered as massless, alongside all quarks (apart from the top quark), as their masses are negligible at these energy scales.
In certain cases, the mass of the bottom quark is also included in theoretical calculations, but with the width assumed to be zero.} as these constitute one of the building blocks of the method. 
As discussed briefly in the introduction, state preparation of probability distributions for QMCI is an active research topic, and several methods have been proposed in the literature (see, \emph{e.g.}  Refs.~\cite{plesch2011quantum,Sanders_2019,zoufal2019quantum,zhang2022quantum,bausch2022fast,rattew2022preparing,wodecki2024spectral}); however, there is currently no standard methodology that is commonly used. 

In this section, we propose two different methods for preparing BW distributions, and discuss the benefits and drawbacks of each. We note that this is by no means meant to be a definitive nor exhaustive list, as the purpose of this section is not to demonstrate the \emph{best possible} solution for preparing BW distributions, but rather to give some examples to illustrate that this can be done in different ways. Indeed, in the future, as new state-preparation methods are developed, these can be utilised to generate larger-scale circuits, improve on the accuracy of the generated distributions, and reduce the quantum resources required to prepare them.

Before we continue the discussion, however, we must first discuss some of the challenges inherent to state preparation for QMCI. These challenges generally arise due to the limited number of qubits, $n$, that are available (due to the limitations of current hardware) to represent the true continuous distribution as a discretised, truncated distribution. In the QMCI engine, the amplitudes of the state, $\ket{p}$, are interpreted as the support points of the discretised distribution, with the number of support points, $N = 2^{n}$. These are uniformly spaced with spacing size, $\Delta$, spanning the range of the support of the distribution, $[x_l,x_u]$. In practice, the limited number of points alongside the uniform binning leads to limited resolution on the distribution of interest, as well as having to truncate what could be an infinite distribution. For a finite number of qubits, these limitations introduce a number of systematic errors into a QMCI calculation.\footnote{It is important to note that many of these systematic errors are also applicable to classical MCI, as will be touched upon.} Ref.~\cite{cui2024uik} discusses these systematic errors in detail, and so here we merely reproduce some of the definitions of the relevant ones for this discussion:
\begin{itemize}
\item \textbf{Discretisation error}: this arises from the finite resolution of sampling for the random variable, due to the limited number of discrete support points used to encode the continuous random variable. This is negligible for classical MCI with (effectively) unlimited numbers of bits, but significant for QMCI due to the limited number of qubits available, therefore

\begin{equation}\label{Eq:integrationToRiemannSum}
    \epsilon_{d} = \abs{\int_{x_l}^{x_u}g(x)f_{X}(x)\,dx - \sum_{i=0}^{N-1}g(x_{i})f_{X}(x_{i})\Delta},
\end{equation}
where $x_{0} = x_{l}$, $x_{i} = x_{l} + i\Delta$, $x_{N-1} = x_{u}$, and $\Delta = x_{i+1} - x_{i} = \cfrac{x_u-x_l}{N}$.

\item \textbf{Normalisation error}:
this arises because truncation and discretisation distort the probability distribution, meaning the mass of the PDF that is loaded as a quantum state may not be equal to unity, therefore
\begin{equation}
    \epsilon_{n} = 
        \sum_{i=0}^{N-1}\abs{g(x_{i})\left(f_{X}(x_{i})\Delta - \tilde{f}_{X}(x_{i})\right)}
    ,
\end{equation}
where $\tilde{f}_{X}(x_{i})$ are the true normalised probabilities.
\item \textbf{Thresholding error}: this occurs specifically for the QMCI engine when considering the thresholding operations used to impose limits of integration. To estimate $\mathbb{E}\left[X\,\Theta(X \geq V_{Th})\right]$, the QMCI engine defines a threshold with an inclusive upper bound, but if $V_{Th} \in \left(x_i, x_{i+1}\right)$, it instead approximates the threshold using $x_i$, therefore
\begin{equation}
    \epsilon_{th} = \abs{
        \mathbb{E}\left[X\,\Theta(X \geq V_{Th})\right] - \mathbb{E}\left[X\,\Theta(X \geq x_{i})\right]
    }.
\end{equation}

\item \textbf{State-preparation error}:
this error is distinct from the other errors discussed as it does not arise from the limited number of qubits, but rather the imperfect method of preparing the distribution as a quantum state \emph{i.e.}\ there is a difference between the ideal quantum state and the approximated state that is actually prepared. The state-preparation error can be defined in a variety of different ways. In this article, we consider the \emph{cumulative-distribution-function} (CDF) \emph{mean-squared error} (CMSE) between the probabilities of the prepared state, $f^{p}_{X}(x_i)$, and the true probabilities, $\tilde{f}_{X}(x_{i})$, as the metric representing the state-preparation error
\begin{equation}
    \epsilon_{s}^{\text{CMSE}} = \frac{1}{N} \sum_{i=1}^N \left( \tilde{F}_X(x_i) - F^{s}_{X}(x_i)\right)^2,
\end{equation} 
where $\tilde{F}_X, F^{s}_{X}$ are the empirical CDFs of $\tilde{f}_{X}$ and  $f^{s}_{X}$, respectively. In addition, we also consider the \emph{Jensen-Shannon divergence} (JSD) as a metric for determining how accurately the state is prepared \begin{equation}
    \text{JSD} = \frac{1}{2} \left(\sum_{i} \tilde{f}_{X}(x_{i}) \log \frac{\tilde{f}_{X}(x_{i})}{\mathcal{F}(x_i)} + \sum_{i} f^{s}_{X}(x_i) \log \frac{f^{s}_{X}(x_i)}{\mathcal{F}(x_i)}\right),
\end{equation} 
where $\mathcal{F}(x_i) \equiv \frac{1}{2}(\tilde{f}_{X}(x_{i})+f^{s}_{X}(x_i))$.
\end{itemize} All of these systematic errors can be considered relevant for the cross-section calculations considered later in \cref{sec:applications}. 

Considering the discretisation and normalisation errors in particular, for a given number of qubits, $n$, there will be a sub-range of the full potential support of the distribution where these systematic errors are sufficiently minimised (below a chosen bound) to have a negligible impact on a QMCI calculation. When considering cross-section calculations for a given experiment, the integration is only ever performed from $ s_{min} =  0 \GeV^{2} $ to some $ s_{max} = S \GeV^{2}$. Therefore, in practice, it will not be necessary to prepare the BW distribution to span the full potential support range [$s_{min}=0 \GeV^{2} $ to $s_{max}=+\infty \GeV^{2}$]---which would require an infinite number of qubits in order to sufficiently suppress these errors. Thus, whilst qubit numbers are still limited, for practical applications it will instead suffice to generate a range of circuits that prepare BW distributions with supports spanning a range of different COM energies squared, with each prepared using a sufficient number of qubits as to sufficiently suppress these systematic errors. 

It is also worth mentioning that while hadron colliders collide particles at energies of up to tens of TeV ($13.6\TeV$ for the LHC), 
due to the suppression of the parton distribution functions at high energies, 
the accessible energy range is effectively reduced to a few hundreds of $\GeV$.

To this end, and for illustration purposes, in this article we consider preparing BW distributions for the $\PW$, $\PZ$, and $\Pt$ resonances, corresponding to a COM energy, $\COM = 200 \GeV$, 
and for just the $\PW$ to $\COM = 100 \GeV$. The value, $\COM = 100\GeV$, was chosen such that the circuits for the cross-section calculations in \cref{sec:separabletwodimensionalintegration,sec:fulltwodimensionalintegration} (which only use the $\PW$-boson propagator) were sufficiently small to be classically simulable using state-vector methods, while $\COM = 200\GeV$ was chosen in order to be able to generate distributions covering all of the resonances.

\begin{figure}
    \centering     \includegraphics[width=0.55\linewidth]{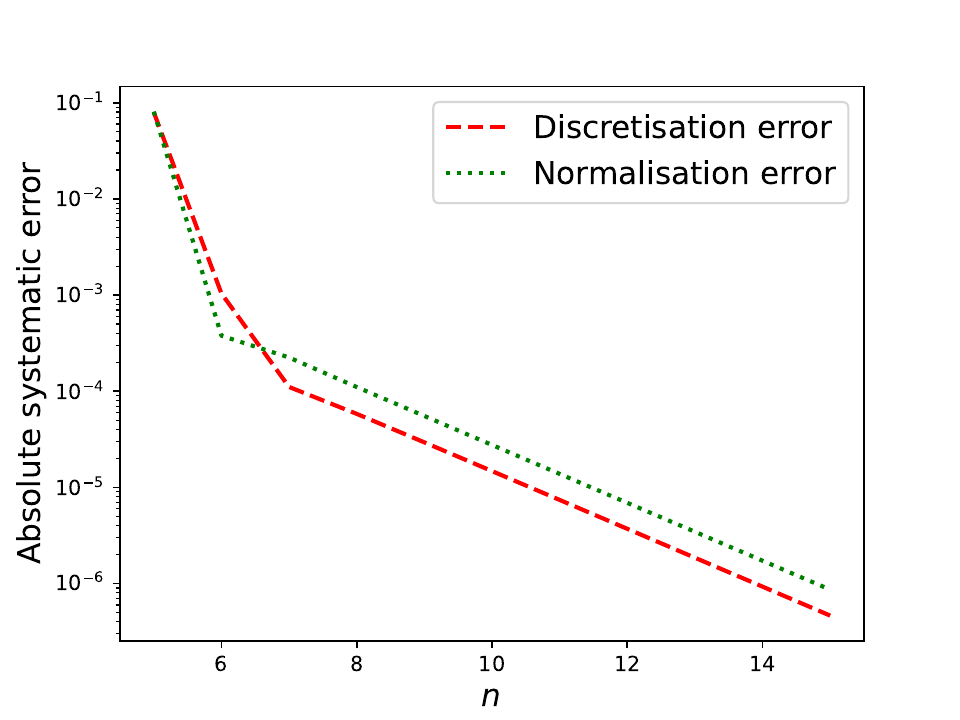}
    \caption{Absolute discretisation and normalisation errors as a function of the number of qubits, $n$, used to prepare the $\PW$-boson BW distribution corresponding to a COM energy $\COM = 100 \GeV$.}
    \label{fig:errors100}
\end{figure}

\begin{figure}
    \centering
    \begin{subfigure}{0.49\textwidth}
        \centering        \includegraphics[width=\linewidth]{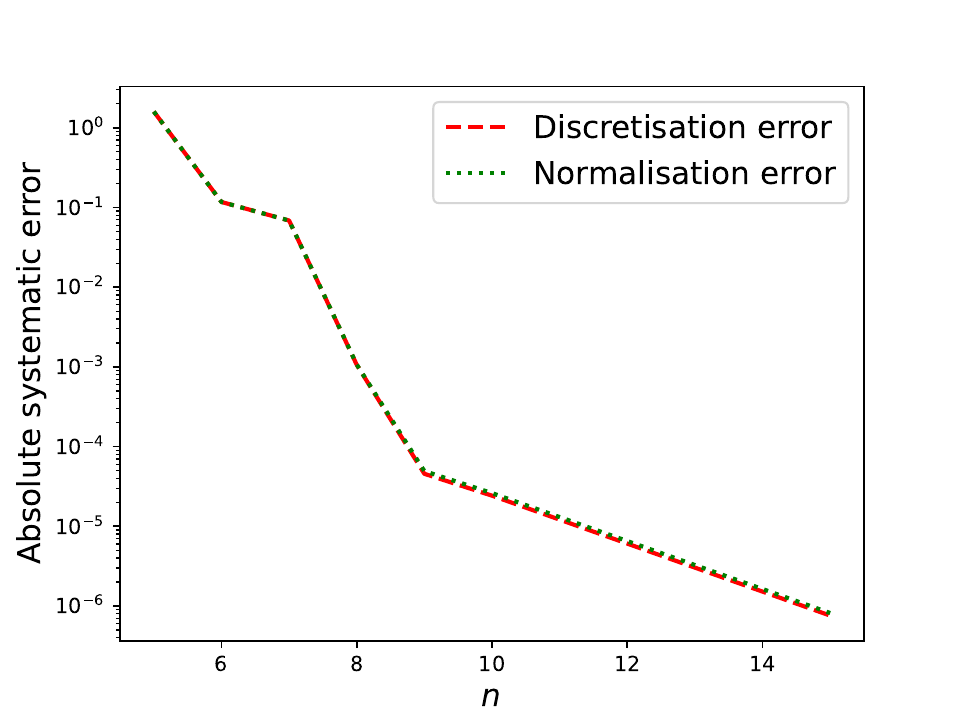}
        (a) $\PW$ boson
    \end{subfigure}
    \begin{subfigure}{0.49\textwidth}
        \centering        \includegraphics[width=\linewidth]{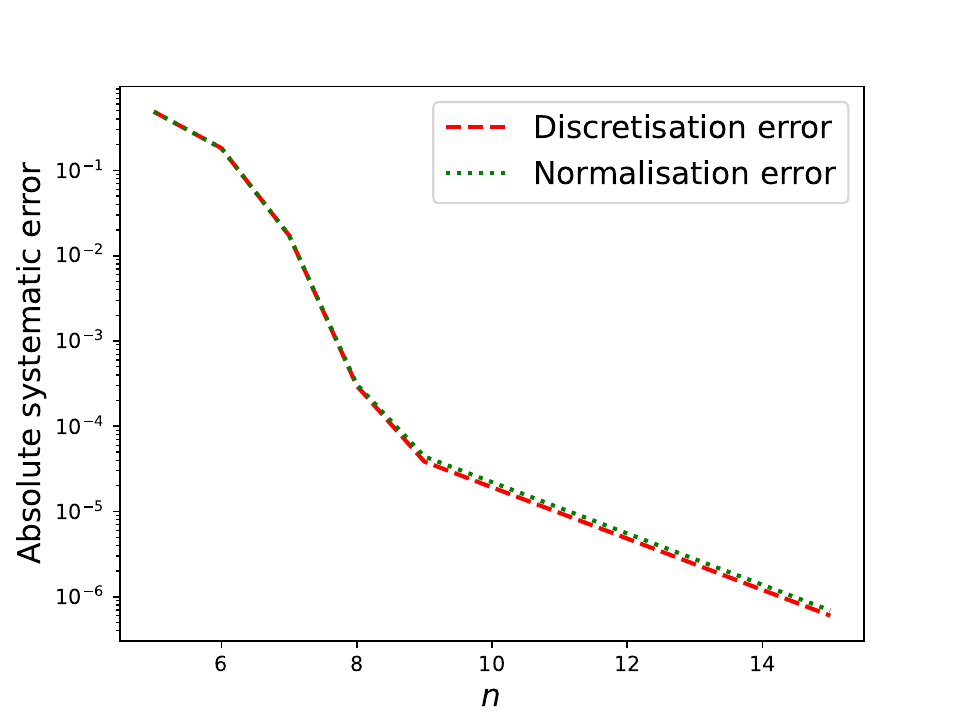}
        (b) $\PZ$ boson
    \end{subfigure}
    \begin{subfigure}{0.49\textwidth}
        \centering        \includegraphics[width=\linewidth]{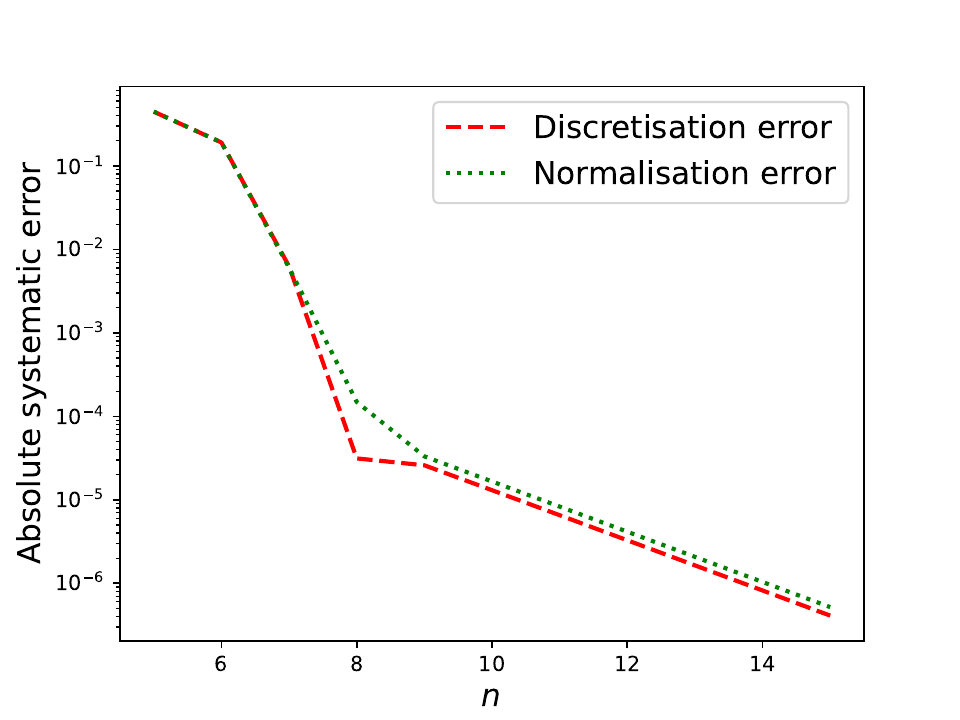}
        (c) $\Pt$ quark
    \end{subfigure}
    \caption{Absolute discretisation and normalisation errors as a function of the number of qubits, $n$, used to prepare the BW distributions for various resonances corresponding to a COM energy $\COM = 200 \GeV$.}
    \label{fig:errors200}
\end{figure}

\Cref{fig:errors100,fig:errors200} plot the discretisation and normalisation errors as a function of the number of qubits used to prepare the distributions for the various propagators. Considering the $\COM = 100 \GeV$ distribution first, with the aim of allowing for classical simulations later, while keeping systematic errors relatively suppressed (\emph{i.e.}\ minimising the number of qubits in the state-preparation circuit), we selected $n=6$, based on \cref{fig:errors100}, which corresponds to errors in the range $\sim (10^{-3} - 10^{-4})$. For the $\COM = 200 \GeV$ distributions we chose $n=9$, such that errors are in the $\sim(10^{-4}-10^{-5})$ range.

We now move on to discuss the two different methodologies investigated for preparing BW distributions on quantum computers.

\subsection{Variational method}
The first methodology is based on a variational, quantum machine-learning approach, as motivated by Ref.~\cite{Akhalwaya:2023hqe}. Here a PQC~\cite{benedetti:2019inj} is trained to generate an approximation of the distribution of interest.

Briefly summarising the method, the loss function that is minimised in the training using a classical optimiser is the $L_{\alpha}$ norm (for $\alpha=1,2,\infty$) between the target state, $\ket{\text{tar}}$, and the generated state, $\ket{\text{gen}}$, \begin{equation}
     ||\ket{\text{tar}} - \ket{\text{gen}}||_\alpha = \left( \sum_{i=1}^N  \left( \left[\ket{\text{tar}} - \ket{\text{gen}} \right]_i \right)^\alpha \right)^{1/\alpha}.
 \end{equation}
 The PQC ansatz chosen, $U(\vec{\theta})$, is a maximally expressive hardware-efficient ansatz \cite{Kandala:2017vok} consisting of an $n$-qubit circuit formed of $L+1$ layers, with all angles initialised to $\pi/2$, giving $n(L+1)$ variational parameters as
 \begin{equation}
U(\vec{\theta}) =  U_R(\vec{\theta}^{ L+1})~ \overbrace{
U_{CX} U_R(\vec{\theta}^{ L}) \ldots U_{CX} U_R(\vec{\theta}^{ 1})}^{L\rm{-times}},
\label{eq:trial_ansatz_u_theta}
\end{equation} where $U_{CX}$ are fixed blocks of CX gates. The training parameters to optimise are the angles of the $R_y$ rotations across $k$ layers
\begin{equation}
U_R (\vec{\theta}^k) = \bigotimes_{i=0}^{n-1} R_y({\theta}_{i}^k).
\end{equation}
Thus the aim is to learn a $U(\vec{\theta})$ such that
\begin{equation}
\ket{\text{gen}} = \ket{\psi(\vec{\theta})} = U(\vec{\theta}) \ket{0^n}.
\label{eq:trial_ansatz}
\end{equation} 

As is well understood from the literature (see, \emph{e.g.}\ Refs.~\cite{mcclean18,cerezo21,sharma22, wang21, marrero20, arrasmith20}), variational methods are plagued by issues of trainability, specifically the gradients of the cost function vanishing exponentially in the size of the system (number of qubits), known as \emph{barren plateaus}. Barren plateaus have been shown to be directly related to the expressitivity of the circuit ans{\"a}tze used in the training \cite{Holmes:2021qjw}. In general, there is therefore a balance between allowing the circuit to be expressive enough to contain the desired solution, whilst also limiting the effects of barren plateaus. Given this, it is thought that variational approaches for state preparation will likely not be scalable for large system sizes. However, for the BW distributions considered in this article---which are all small-scale circuits---variational methods should give good results in practice. Indeed, no particular effort is made here to limit the expressivity of the circuit ansatz (for example by using some alternative ans{\"a}tze such as the ones discussed in Refs.~\cite{grant19, volkoff21}), as it was not considered necessary for these small systems.

\subsection{Fourier expansion method}
The second methodology that we investigate for preparing BW distributions involves a promising recent technique by Rosenkranz \emph{et al.}~\cite{BenedettiRosenkranzStatePrep2024}, where a multivariate probability distribution is decomposed into a Fourier series. In our case we need only consider univariate state preparation.

To again briefly summarise the method, the target function,$f$, is approximated by mapping to the interval $[-1,1]$\footnote{Specifically, it is represented in the interval $[0,1]$ and then a periodic extension is applied to map onto the interval $[-1,0)$.} with a finite Fourier series of the form 
\begin{equation}
f_{d}(x) = \sum^d_{k=-d}c_{k}e^{i\pi kx},
\end{equation}
where $d$ is the degree of the expansion and $c_{k}$ are the Fourier coefficients. An interpolation is used such that $f_d$ matches $f$ for a number of interpolation points, and the coefficients $c_{k}$ for the interpolant are calculated using the fast Fourier transform. The basis functions of the Fourier expansion up to a chosen $d$ are then prepared efficiently on a quantum register using a block encoding. The weighted sum of basis functions is prepared using a \emph{linear combination of unitary} (LCU) operations \cite{childs2012gwh}, leading to a block-encoding circuit for $f_{d}$. 

The potential benefits of this method is that, in contrast to the previous method, it should scale well to larger system sizes. However, one important consideration is that the method relies on a considerable number of ancilla qubits, and is a probabilistic state preparation; this is because the LCU method requires $\text{log}_{2}(2d+1)$ additional ancilla qubits, and to then prepare the correct state, all of these ancilla qubits must be post selected to give the zero state, which has success probability $p_{\text{success}}=\frac{\sum_x \abs{f_{d}(x)}^2}{2^{n}(\sum_k\abs{c_k})^2}$. In practice, $p_{\text{success}}$ can be increased using amplitude amplification, although this of course means additional quantum resources. It is also worth noting that, in practice, the chosen value of $d$ dictates the final accuracy of the prepared distribution (\emph{i.e.}\ larger $d$ means a smaller state-preparation error and JSD).

\subsection{Results and comparison}

Both the methods discussed in the previous sections were used in order to produce circuits that prepared BW distributions for the $\PW$, $\PZ$, and $\Pt$ resonances based on the $\COM$ ranges and corresponding numbers of qubits in the circuit ansatz. For comparing the different methods, the metrics that are considered are the resources required for the circuit, in terms of the total number of qubits, $n$, alongside the number of one- and two-qubit gates, $g_{1q}$ and $g_{2q}$, respectively, and the accuracy of the prepared distribution, in terms of the state-preparation error, $\epsilon_{s}^{\text{CMSE}}$, and the JSD. In this section we will explicitly discuss the nine-qubit circuits corresponding to $\COM=200 \GeV$, in order to perform a comparison of the two methods for all resonances. However, results for the other COM energy range are also briefly discussed later.

PQCs were trained based on the variational method, as discussed previously. In all cases, the metric used in the cost function was the $L_2$ norm between the target and generated state. The \emph{basin-hopping} algorithm \cite{olson12} implementing the \emph{Broyden–Fletcher–Goldfarb–Shanno} method \cite{nocedal06} was used for the minimisation. A coarse hyperparameter search was performed during training in order to optimise the performance, based on varying the number of layers (parameters) in the ans{\"a}tze alongside another hyperparameter related to the cost function (the overall scaling factor).

The Fourier expansion method was then analysed. For each of the resonances we produced a circuit that was roughly equivalent to the best one produced by the variational method in terms of the accuracy of the prepared distribution. In addition, we also produced another circuit where the accuracy of the prepared distribution was significantly greater than in the variational case (\emph{i.e.}\ with an increased value of $d$). This was done to demonstrate that the method enables the preparation of larger-scale circuits while maintaining accuracy.

\begin{figure}
    \centering
    \begin{subfigure}{0.49\textwidth}
        \centering
        \includegraphics[width=\linewidth]{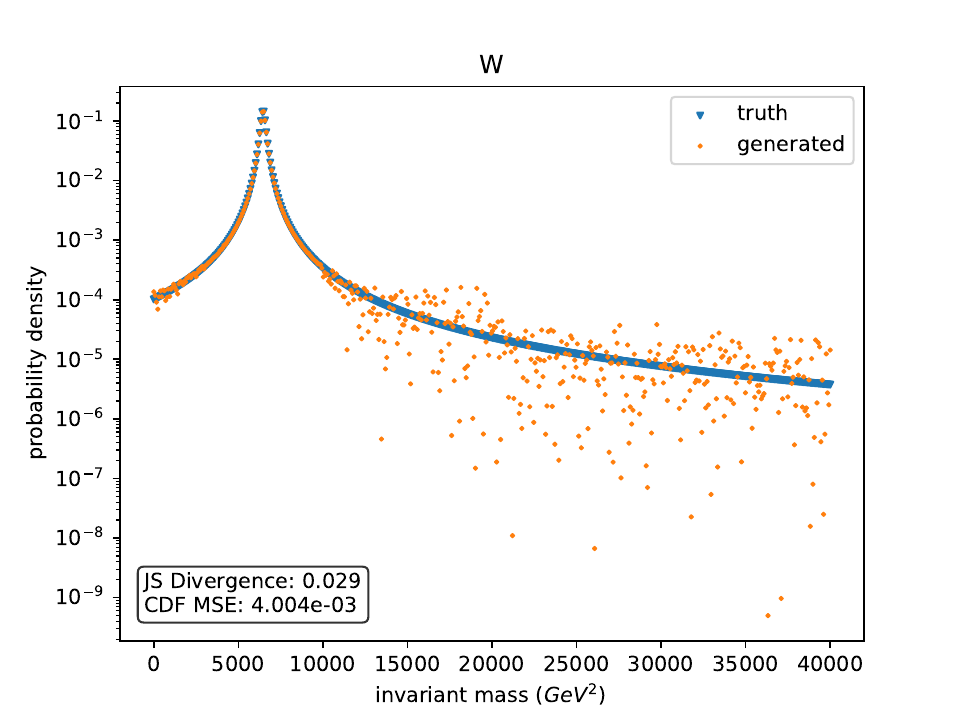}
        (a) $\PW$ boson
    \end{subfigure}
    \begin{subfigure}{0.49\textwidth}
        \centering
        \includegraphics[width=\linewidth]{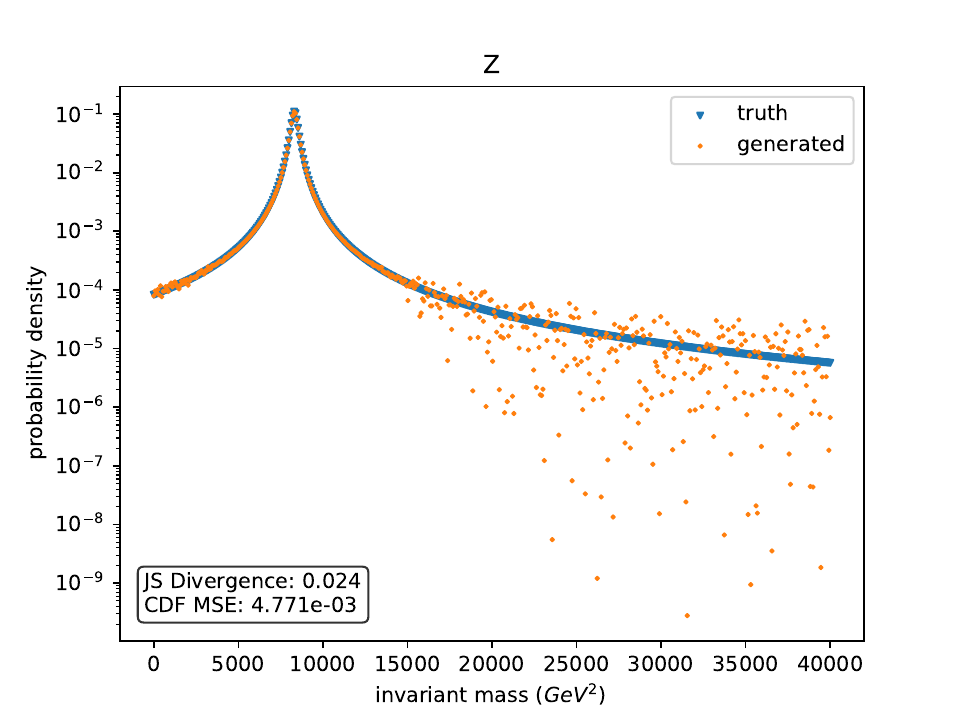}
        (b) $\PZ$ boson
    \end{subfigure}
    \begin{subfigure}{0.49\textwidth}
        \centering
        \includegraphics[width=\linewidth]{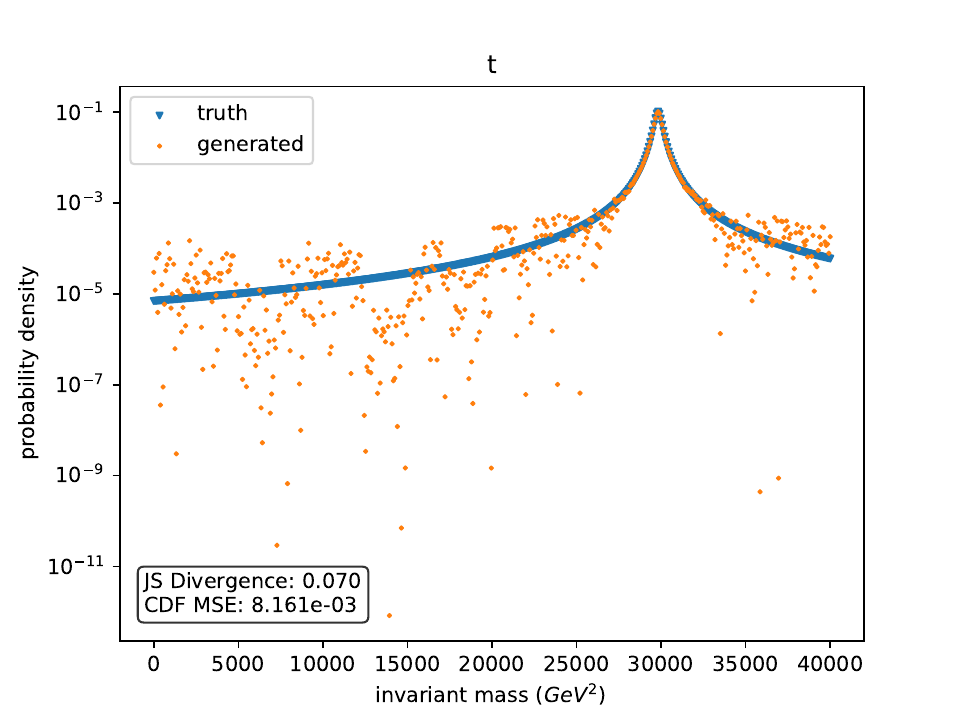}
        (c) $\Pt$ quark
    \end{subfigure}
    \caption{True (blue triangles) and generated (orange circles) points for nine-qubit circuits for the BW distribution for various resonances up to $\COM=200 \GeV$, generated using the variational method.}
    \label{fig:distcomparisons9q_var}
\end{figure}

\begin{figure}
    \centering
    \begin{subfigure}{0.49\textwidth}
        \centering \includegraphics[width=0.95\linewidth]{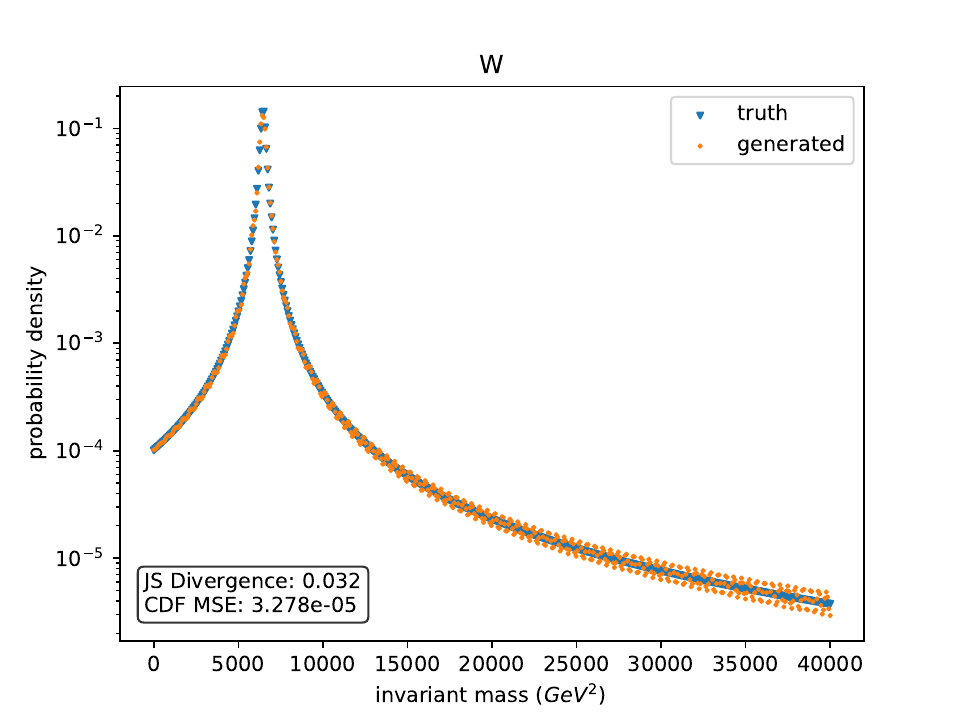}
        (a) $d=175$
    \end{subfigure}
    \begin{subfigure}{0.49\textwidth}
        \centering \includegraphics[width=0.95\linewidth]{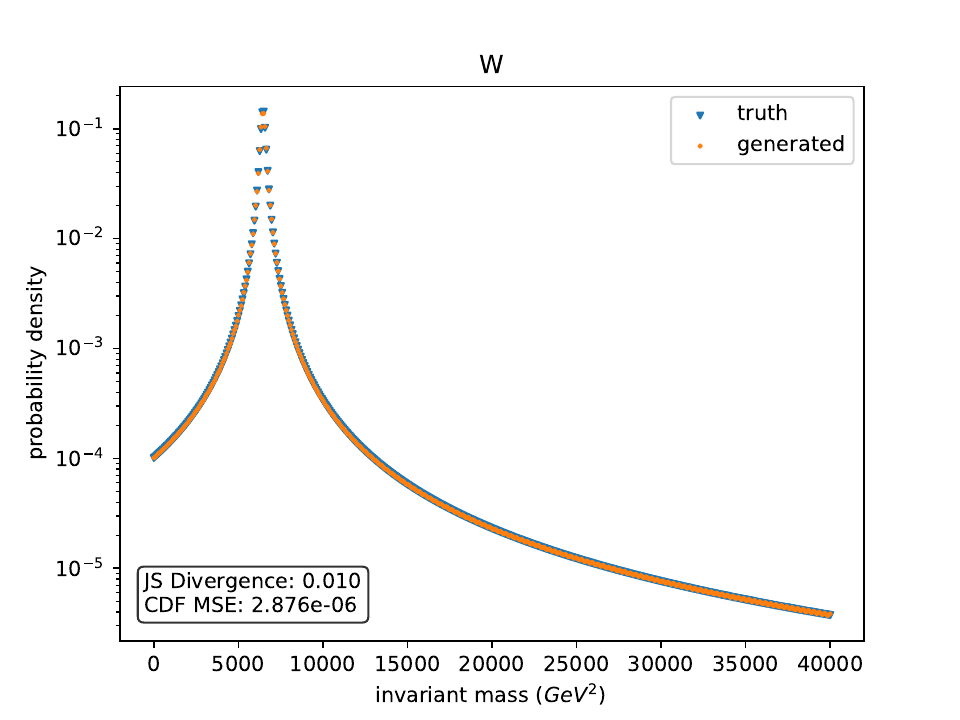}
        (b) $d=250$
    \end{subfigure}
    \caption{True (blue triangles) and generated (orange circles) points for the $\PW$-boson BW distribution up to $\COM=200 \GeV$, generated using the Fourier expansion method, for two different nine-qubit circuits corresponding to (left) a less accurate generated distribution that matches the accuracy of the variational method, and (right) a more accurate generated distribution.}
    \label{fig:w_lcu}
\end{figure}

\begin{figure}
    \centering
    \begin{subfigure}{0.49\textwidth}
        \centering      \includegraphics[width=0.95\linewidth]{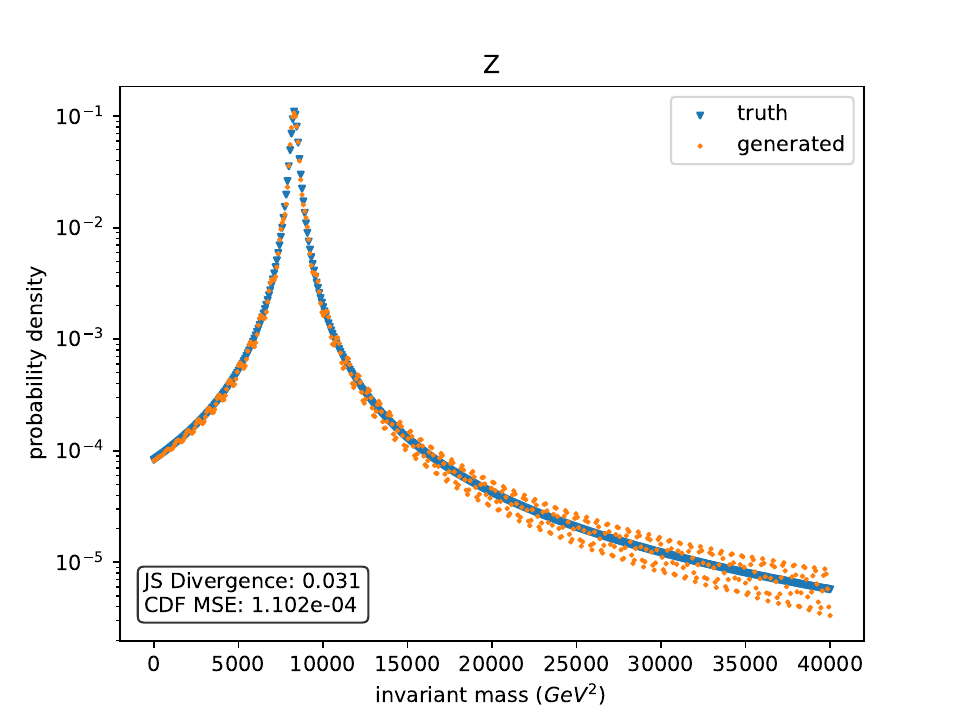}
        (a) $d=130$
    \end{subfigure}
    \begin{subfigure}{0.49\textwidth}
        \centering      \includegraphics[width=0.95\linewidth]{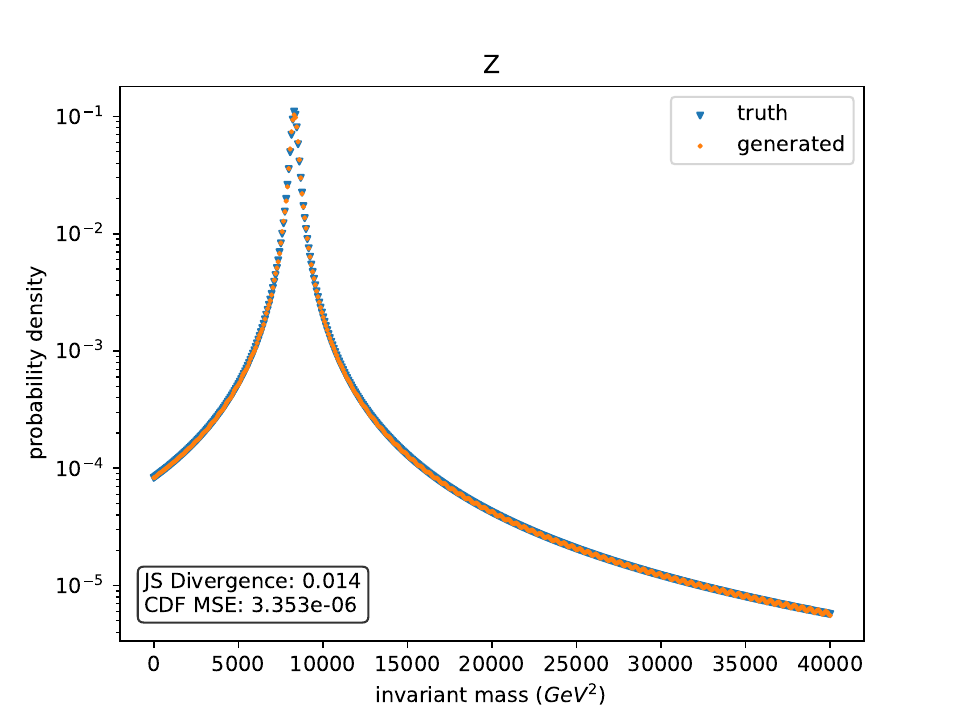}
        (b) $d=170$
    \end{subfigure}
    \caption{True (blue triangles) and generated (orange circles) points for the $\PZ$-boson BW distribution up to $\COM=200 \GeV$, generated using the Fourier expansion method, for two different nine-qubit circuits corresponding to (left) a less accurate generated distribution that matches the accuracy of the variational method, and (right) a more accurate generated distribution.}
    \label{fig:z_lcu}
\end{figure}

\begin{figure}
    \centering
    \begin{subfigure}{0.49\textwidth}
        \centering      \includegraphics[width=0.95\linewidth]{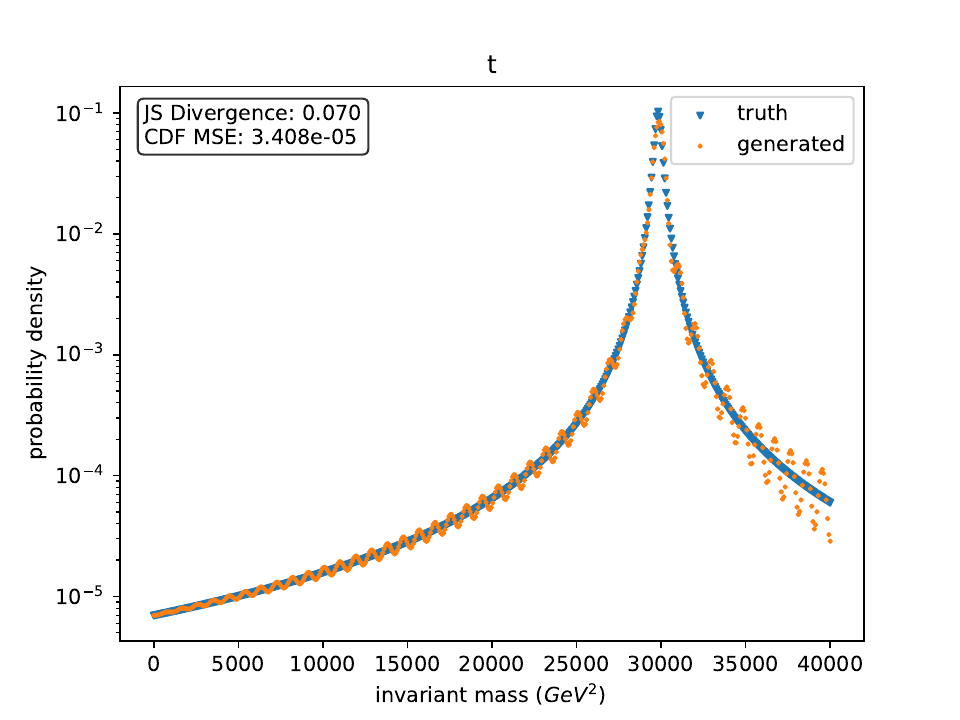}
        (a) $d=85$
    \end{subfigure}
    \begin{subfigure}{0.49\textwidth}
        \centering      \includegraphics[width=0.95\linewidth]{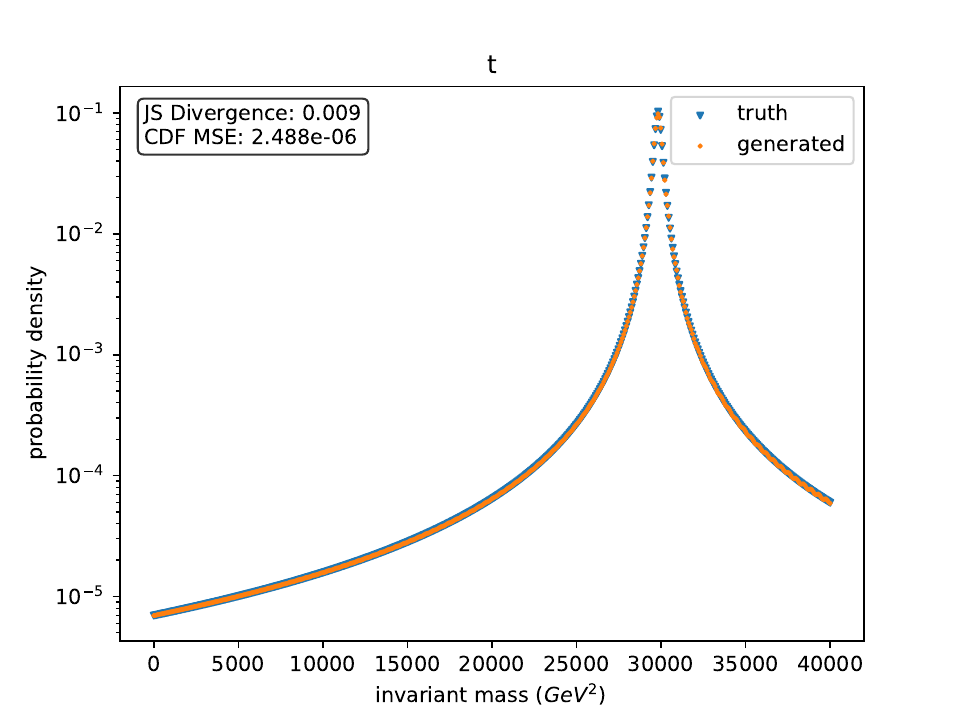}
        (b) $d=180$
    \end{subfigure}
    \caption{True (blue triangles) and generated (orange circles) points for the $\Pt$-quark BW distribution up to $\COM=200 \GeV$, generated using the Fourier expansion method, for two different nine-qubit circuits corresponding to (left) a less accurate generated distribution and (right) a more accurate generated distribution.}
    \label{fig:t_lcu}
\end{figure}

\Cref{fig:distcomparisons9q_var} gives the generated distributions for each resonance using the variational method, while \cref{fig:w_lcu,fig:z_lcu,fig:t_lcu} give the generated distributions using the Fourier method for each of the resonances, respectively. \Cref{tab:bw_distributions} lists the metrics for the distributions produced for all COM ranges for both methods. \Cref{fig:w_100GeV} in \cref{app:stateprep} also plots the distributions generated by the variational and Fourier expansion methods for the $n=6$, $\COM=100\GeV$ circuits.

\begin{table}
\centering
\caption{Comparison of the metrics for the prepared distributions for each resonance for both COM energies, using either variational or Fourier expansion methods. `Res' stands for `Resonance'. For `Accuracy', `Optimised' refers to the variationally-trained circuits with optimised hyperparameters, and `Matched' and `More' refer to the Fourier-expansion-method circuits, with either similar, or much greater accuracy, to the equivalent variationally trained circuits, respectively.}
\resizebox{\textwidth}{!}{%
\begin{tabular}{llclcccccc}
\toprule
\textbf{$\COM$} & \textbf{Method} & \textbf{Res} & \textbf{Accuracy} & \textbf{$n$} & \textbf{$g_{1q}$} & \textbf{$g_{2q}$} & \textbf{$\epsilon_{s}^{\text{CMSE}}$} & \textbf{JSD} & $p_{\text{success}}$ \\
\midrule
\multirow{3}{*}{$100 \GeV$} 
& Variational & $\PW$ & Optimised & $6$ & $186$ & $150$ & $1.22\times10^{-4}$ & $3.31\times10^{-6}$ & N/A \\
\cmidrule{2-10}
& Fourier & $\PW$ & Matched ($d=250$) & $15$ & $1592$ & $1638$ & $1.22\times10^{-4}$ & $6.73\times10^{-5}$ & $3.12\%$\\
\midrule
\multirow{10}{*}{$200 \GeV$} 
& \multirow{3}{*}{Variational} 
& $\PW$ & Optimised & $9$ & $180$ & $152$ & $4.01\times10^{-4}$ & $0.029$ & N/A \\
&& $\PZ$ & Optimised & $9$ & $234$ & $200$ & $4.77\times10^{-3}$ & $0.024$ & N/A \\
&& $\Pt$ & Optimised & $9$ & $126$ & $104$ & $8.16\times10^{-3}$ & $0.070$ & N/A \\
\cmidrule{2-10}
& \multirow{6}{*}{Fourier} 
& \multirow{2}{*}{$\PW$} & Matched ($d=175$) & $18$ & $1576$ & $1684$ & $3.28\times10^{-5}$ & $0.032$ & $0.9\%$ \\
&& & More ($d=250$) & $18$ & $1602$ & $1686$ & $2.13\times10^{-6}$ & $0.003$ & $0.9\%$ \\
&& \multirow{2}{*}{$\PZ$} & Matched ($d=130$) & $18$ & $1591$ & $1692$ & $3.28\times10^{-5}$ & $0.031$ & $1.1\%$ \\
&& & More ($d=170$) & $18$ & $1601$ & $1686$ & $2.58\times10^{-6}$ & $0.014$ & $1.2\%$ \\
&& \multirow{2}{*}{$\Pt$} & Matched ($d=85$) & $17$ & $825$ & $906$ & $3.41\times10^{-5}$ & $0.070$ & $1.4\%$ \\
&& & More ($d=180$) & $18$ & $1593$ & $1692$ & $2.45 \times 10^{-6}$ & $0.010$ & $1.2\%$ \\
\bottomrule
\end{tabular}
}
\label{tab:bw_distributions}
\end{table}

In order to compare the two approaches, we will only discuss the $\PW$-boson results, as the same arguments apply to the others. We note that the Fourier expansion method allows for much more accurate states to be prepared than the variational method, and as mentioned previously, the real benefit is that it is scalable to larger numbers of qubits. However, we found that the resources required were significantly larger when considering the equivalent accuracy circuits to those produced using the variational method, due to the large number of additional ancilla qubits required. In addition, because the Fourier expansion method is a probabilistic state-preparation method, then in practice, without using techniques such as amplitude amplification, such small success probabilities will require the circuit to be run several times before the desired state is actually prepared (successful post selection).

It is worth also discussing that, for the $n=6$ circuits (with distributions plotted in \cref{fig:w_100GeV} in \cref{app:stateprep}), we find that the variational method produces slightly more accurate results than the Fourier expansion method, and again requires significantly less resources. This indeed makes sense, given that we expect variational methods to perform well for small-scale circuits; however, one would expect the performance to decrease as the system size increases, which is indeed what we see when considering the $n=9$ circuits discussed previously.

\begin{figure}[ht!]
    \centering
    \begin{subfigure}{0.49\textwidth}
        \centering        \includegraphics[width=\linewidth]{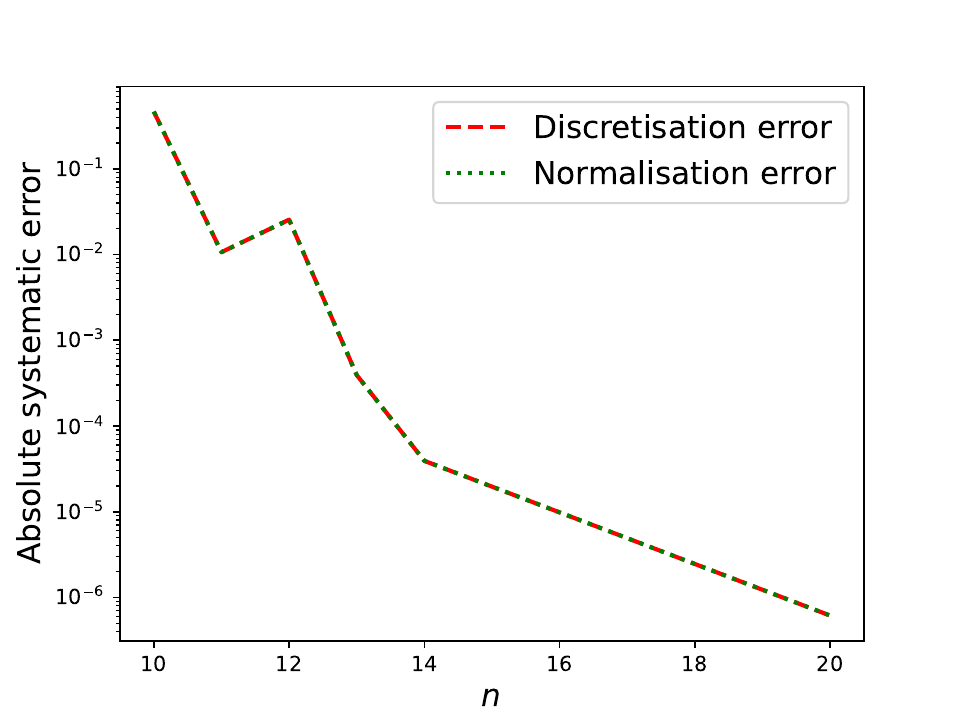}
        (a) $\PW$ boson
   \end{subfigure}
    \begin{subfigure}{0.49\textwidth}
        \centering        \includegraphics[width=\linewidth]{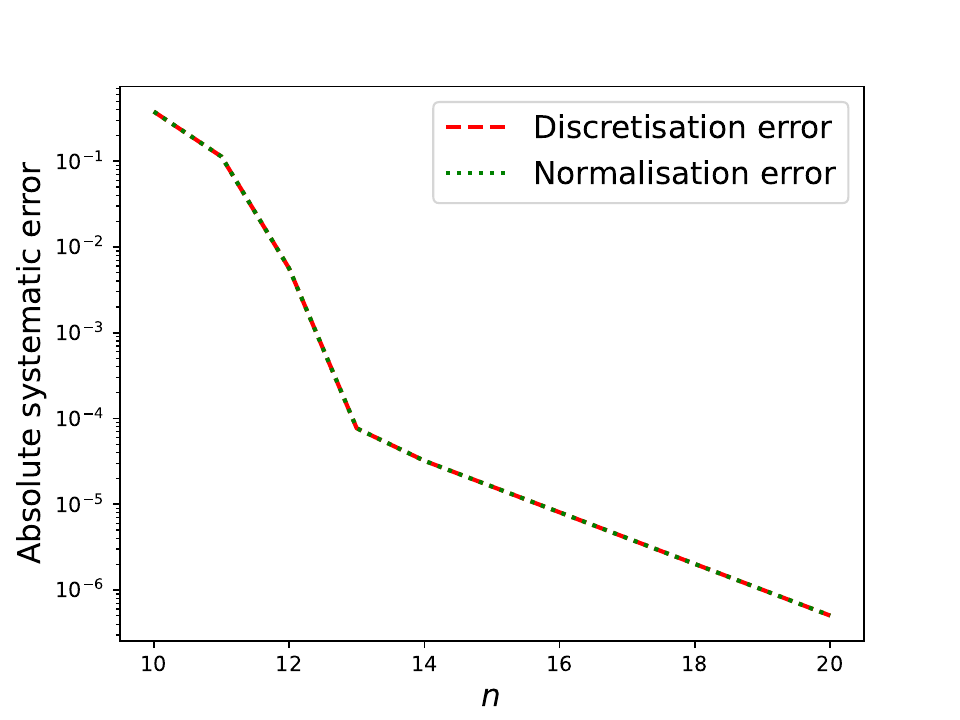}
        (b) $\PZ$ boson
    \end{subfigure}
    \begin{subfigure}{0.49\textwidth}
        \centering        \includegraphics[width=\linewidth]{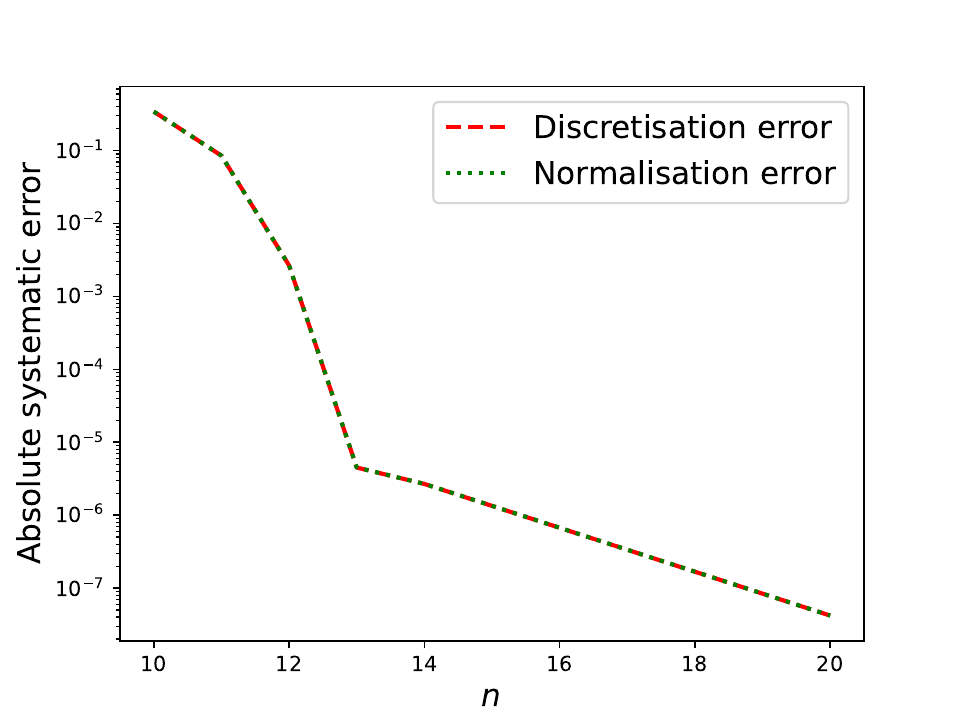}
        (c) $\Pt$ quark
    \end{subfigure}
    \caption{Absolute discretisation and normalisation errors as a function of the number of qubits, $n$, used to prepare the BW distributions for various resonances corresponding to a COM energy, $\COM = 1 \TeV$.}
    \label{fig:errors1000}
\end{figure}

As an aside, if we were to consider an example of a potential more realistic use case corresponding to an energy scale comparable to a modern collider experiment---where in reality it would be essential to really suppress systematic errors---of $\COM = 1\TeV$, for example, then we can see from \cref{fig:errors1000} that we would need around $n=16$ for systematic errors to be suppressed to around $\sim(10^{-5}-10^{-6})$. For circuits this size, variational methods are likely not to be performant due to the trainability issues previously discussed. On the contrary, the Fourier expansion method should still be effective for generating accurate distributions at this circuit size, although the resources required would likely be considerable.

As discussed previously, state-preparation for probability distributions is an active research topic, and future developments should hopefully allow for the preparation of accurate BW distributions with reduced circuit sizes, paving the way for less resource-intensive QMCI for HEP. We now move on to discuss an example application of the QMCI engine for calculating a cross-section in HEP, making use of the six-qubit, $\COM=100\GeV$, variationally-trained circuit for the $\PW$-boson BW distribution discussed in this section.

\section{Example applications}
\label{sec:applications}
\begin{figure}
    \centering
        \centering        \includegraphics[width=0.49\textwidth]{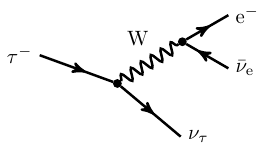}
    \caption{Feynman diagram for decay of a tau fermion [see \cref{eq:tau_decay}].}
    \label{fig:Feynman}
\end{figure}

To illustrate the generality of the approach sketched above, we will focus on one particular example, the decay of the tau lepton into three fermions
\begin{align}
\label{eq:tau_decay}
    \tau^{-} \to{}& \nu_{\tau} \Pe^{-} \bar{\nu_{\Pe}}, \\
    p  ={}& k_1 + k_2 + k_3\nonumber ,
\end{align} 
where $p$ and $k_i$ are the initial and final state four momenta, respectively.
In this case, the matrix element squared can be written as\footnote{The expression has been obtained with the help of the {\sc FeynArts}~\cite{Kublbeck:1990xc, Hahn:2000kx} and {\sc FormCalc}~\cite{Hahn:1998yk} packages to the lowest order in perturbation theory.}
\begin{align}
\label{eq:ME}
    |\mathcal{M}|^2 = - \frac{\alpha^2 \pi^2}{\sin^4 \theta_{\rm w}} \frac{s_{1}^2 - M_\tau^2 s_{1}}{(s_{2}-M_\PW^2)^2 + \Gamma_\PW \MW} ,
\end{align} 
where the invariants, $s_1$ and $s_2$, are equal to $(p_1+p_3)^2$ and $(p_2+p_3)^2$, respectively.
The electroweak coupling is denoted by $\alpha$, while the weak mixing angle is $\theta_{\rm w}$.
The mass and width of the W bosons are denoted by $m_\PW$ and $\Gamma_\PW$, respectively, while $M_\tau$ is the mass of the tau lepton.\footnote{It is worth noting that the expression for the matrix element in \cref{eq:ME} assumes a massless electron, which is a common approximation in HEP. This is because the mass of the electron ($0.5\MeV$) is negligible with respect to a collision energy ($13.6\TeV$ for the LHC, for example).}
It is worth pointing out that the expression in \cref{eq:ME} contains a W-boson propagator term [see \cref{eq:propagator}]. 
This originates from the fact that the electron and anti-electron neutrino are produced through the decay of an intermediate W boson, as depicted in the Feynman diagram in \cref{fig:Feynman}.

For a three-particle phase space, the integration measure reads~\cite{Byckling:1971vca}
\begin{align}
    \rd \Phi_3 ={}& \prod^3_{i=1} \frac{\rd^3 p_i}{2 E_i} \delta^4 (p-k_1-k_2-k_3) \nonumber \\
    ={}& \frac{1}{32 s} \rd s_1 \rd s_2 \rd \Omega_1 \rd \phi_3 ,
\end{align}
where $\Omega_1$ is a solid angle, and $\phi$ a rotation angle.
The integration boundaries are $s_1 \in [0, s-s_2]$ and $s_2 \in [0, s]$, respectively, where $s$ is the available energy of the system, \emph{i.e.}\ $s=M_\tau^2$ for a tau decay.
For what follows, we will discuss this particular example and some modifications of it (either by simplifying it or rendering it more complex).

\subsection{One-dimensional integration}
\label{sec:onedimensionalintegration}
To demonstrate the methodology, for simplicity, we start by considering only the numerator of \cref{eq:ME}.
By neglecting the constant factors and integrating only over the two invariants $s_1$ and $s_1$, one obtains the following integral
\begin{equation}
\sigma  = \int^{s}_{0} \rd s_2 \int^{s-s_2}_{0} \rd s_1  \left(s_1^2-s_1 M_{\tau}^2\right),
\end{equation}
which has analytical value $\sigma =\frac{s^4}{12} - M_{\tau}^2 \frac{s^3}{6}$ 

In order to implement such a calculation using the QCMI engine, noting that the integral over $s_2$ trivially gives $s$, then we encode the calculation as the following one-dimensional integration with a two-dimensional cut function, $C(s_1,s_2)$,
\begin{equation}\label{eq:qmci_one_dim_eq}
\sigma = s \left( \int^{s}_{0} \rd s_1 s_1^2 C(s_1,s_2) - M_{\tau}^2 \int^{s}_{0} \rd s_1 s_1 C(s_1,s_2)  \right),
\end{equation} where 
$C(s_1,s_2) = 1$, if $s_1+s_2 < s$, and $C(s_1,s_2) = 0$, otherwise. If we compare this to the general form of the expectation calculated using the QMCI engine in \cref{eq:functionappliedintergral2D}, and identify $s_1=x$ and $s_2=y$ (this identification will remain for all subsequent examples), we observe that there is freedom in the integrand separation, in terms of what to define as the probability distribution, $f_{S_1S_2}(s_1,s_2)$, and the function applied $g(s_1,s_2)$. In this case, for simplicity, we choose to set $f_{S_1S_2}(s_1,s_2)=U(s_1)U(s_2)$, where $U(.)$ is the uniform distribution, and $g(s_1,s_2)=s_1^{2}$ or $g(s_1,s_2)=s_1$.\footnote{Note that this choice does not leverage the capabilities of the QMCI engine in terms of decomposing the integrand into building blocks, as discussed previously---however, that is not the purpose of this initial example.}

As discussed previously in \cref{sec:qmci_engine}, the QMCI engine contains powerful, in-built functionality for performing such a calculation; the cut function can be straightforwardly implemented using thresholding operations, whilst the engine contains efficient methods for calculating both $g(s_1, s_2)=s_1^{2}$ and $g(s_1, s_2)=s_1$. The QMCI engine can thus be used to build efficient, low depth circuits that give estimates of both expectation values on the RHS of \cref{eq:qmci_one_dim_eq}.

For this example, and the ones described later in \cref{sec:separabletwodimensionalintegration,sec:fulltwodimensionalintegration}, we considered three different levels of expected precision, corresponding to upper bounds on the expected RMSE of the final estimator of the order $10\%$, $1\%$, and $0.1\%$,\footnote{The per-mille precision represents the typical accuracy of classical MCI calculation in HEP.} respectively (not accounting for the systematic errors related to the state preparation discussed in \cref{sec:statepreparation}).
For illustration purposes, we run noiseless simulations of the circuits for the case of $10\%$ precision.

We considered a decay-like process where $\COM = M_{\tau} = 1.776 \GeV$, giving an analytical value $\sigma = -8.248$.\footnote{Given that this quantity is a proxy of a cross section, it is not physical, and thus not necessarily positive.} We set $n=5$ for each of the dimensions, $s_1$ and $s_2$, respectively. This results in a final QMCI circuit containing $23$ qubits in total. The circuits to prepare the uniform distributions used to model the probability distributions for each dimension were loaded directly in the QMCI engine using the in-built state-preparation library, and these then combined to construct the bivariate probability distribution, $f_{S_1S_2}(s_1,s_2)$. 

In order to demonstrate the validity of the approach, we ran noiseless simulations using the QMCI engine to numerically estimate the RHS of \cref{eq:qmci_one_dim_eq}, which we denote as $\hat{\sigma}$, $100$ different times, based on an expected precision of $10\%$. The numerical experiments, and the ones described later in \cref{sec:separabletwodimensionalintegration,sec:fulltwodimensionalintegration}, were carried out using {\sc Qulacs}' noiseless state-vector simulator \cite{Suzuki_2021}. The QAE algorithm used in all cases was an in-built implementation of the \emph{maximum-likelihood QAE} algorithm \cite{suzuki2020amplitude}, that is optimised to be able to run for any given number of samples, thereby maximising performance. 

\begin{figure}
\centering
\includegraphics[width=0.6\linewidth]{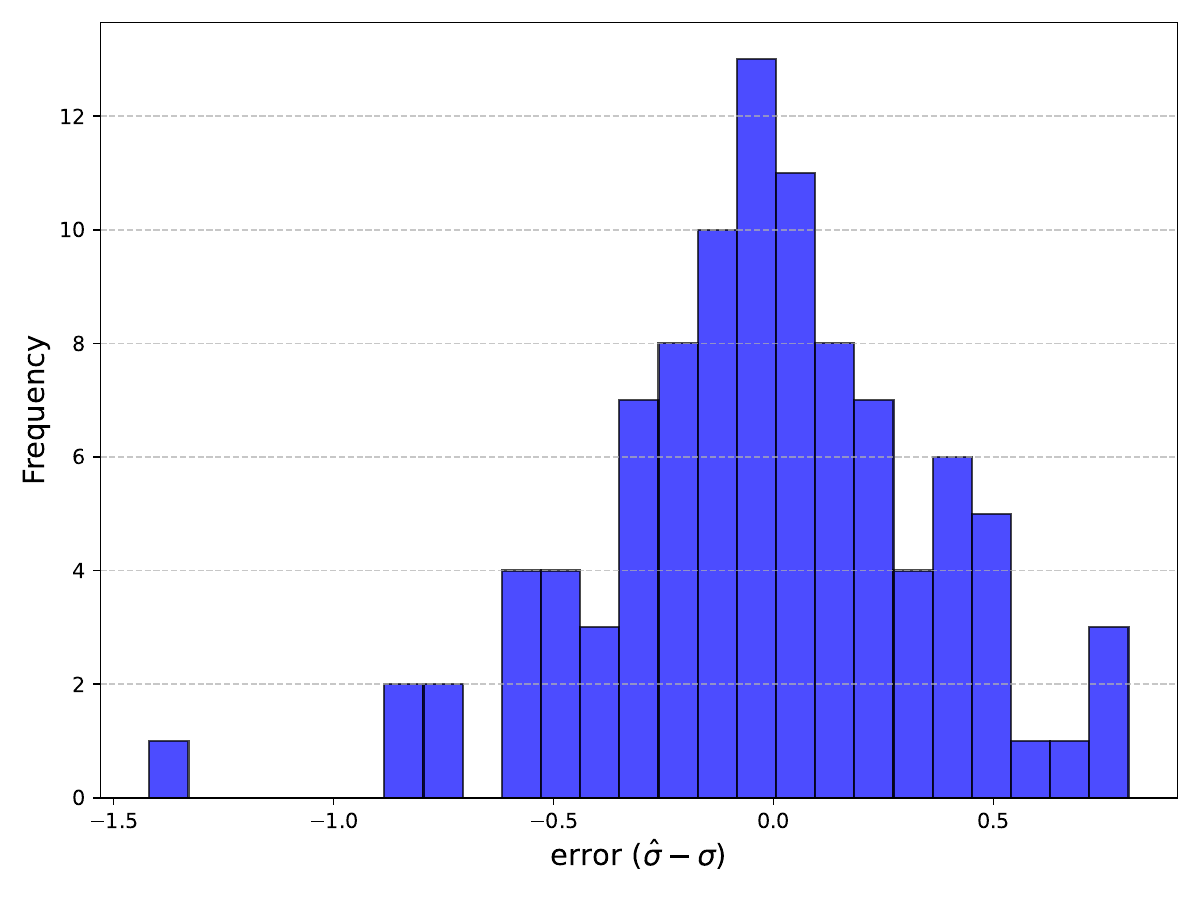}
\caption{Histogram of the error for $100$ runs of QMCI estimating the RHS of \cref{eq:qmci_one_dim_eq}, for an expected RMSE upper-bound precision of $10\%$.}\label{fig:1dintegrationerror}
\end{figure}

\Cref{fig:1dintegrationerror} gives a histogram of the error, $\hat{\sigma} - \sigma$, where we can see that the results are as expected, with nearly all (except for a single outlier) values falling within an expected precision of $10\%$.

Using the in-built resource mode of the QMCI engine (described in \cref{sec:qmci_engine}), we also calculated the resources required to perform this calculation to the three different levels of expected precision. \cref{tab:resources_1Dsimplified} gives these resources for both NISQ and fault-tolerant compilation. We note that the number of qubits required is in general small [$\mathcal{O}(10)$], and in particular is well within the reach of current quantum hardware (this is assuming that the systematic errors based on using $5$-qubit circuits to prepare the uniform distributions do not excessively affect calculations---something that is not observed for this case for $10\%$ precision). However, the number of operations required for NISQ compilation almost certainly means that logical qubits would be required; considering the largest circuit for an expected precision of $10\%$ has $1.80\times10^{6}$ CX gates, then a simple rule-of-thumb calculation suggests one would need a machine with a two-qubit gate fidelity of approximately $1-1/(1.80\times10^{6}) = 99.99994\%$---well out of reach of current quantum hardware. Given the substantial two-qubit gate counts, even the application of targeted error mitigation strategies---such as noise-aware QAE techniques \cite{herbert24}---is unlikely to suffice. There is only so much error mitigation can do in the presence of such high noise levels, and the accumulation of errors is likely to compromise the accuracy of the results, regardless. We note that the number of T-gate operations for fault-tolerant execution are also high; for example, for a $0.1\%$ precision, as would be required to compete with current classical MCI methods, the largest circuit requires $1.71\times10^{9}$ T gates. However, it is worth noting these gate counts are likely to reduce significantly in future as more research into optimised synthesis is carried out (see Section~10.3 of Ref.~\cite{Akhalwaya:2023hqe} for a detailed discussion of this topic), and, therefore, these values should be regarded as loose upper bounds, and treated with a degree of uncertainty.

\begin{table}[ht!]
\centering
\caption{Resources required to estimate the RHS of \cref{eq:qmci_one_dim_eq} using QMCI for various expected precisions, for both NISQ and fault-tolerant resource mode.}
\label{tab:resources_1Dsimplified}
\resizebox{\textwidth}{!}{%
\begin{tabular}{lccccc}
\toprule
\textbf{Compilation}           & \textbf{Resource} & \textbf{Metric}                 &  & \textbf{Precision} &  \\
          &  &                 & \textbf{10\%} & \textbf{1\%} & \textbf{0.1\%} \\
\midrule
                        \textbf{NISQ}    &   Number of qubits            & Largest across circuits        & $24$                      & $24$                      & $24$                      \\
\midrule
                    &      CX gates           & Total number across circuits    & $3.99 \times 10^{6} $             & $4.68 \times 10^{7}$                      & $6.26 \times 10^{8}$                      \\
                            &               & Total depth across circuits     & $2.52 \times 10^{6} $               & $2.94 \times 10^{7}$                      & $3.93 \times 10^{8}$                      \\
                            &               & Number in largest circuit       & $1.80 \times 10^{6} $               & $1.14 \times 10^{7}$                      & $1.69 \times 10^{8}$                      \\
                            &               & Depth of largest circuit        & $1.13 \times 10^{6} $                 & $7.18 \times 10^{6}$                      & $1.06 \times 10^{8}$                      \\
\midrule
                            &    All gates           & Total number across circuits    & $7.97 \times 10^{6} $               & $9.34 \times 10^{7}$                      & $1.25 \times 10^{9}$                      \\
                            &               & Total depth across circuits     & $4.71 \times 10^{6} $               & $5.51 \times 10^{7}$                      & $7.37 \times 10^{8}$                      \\
                            &               & Number in largest circuit       & $3.59 \times 10^{6} $               & $2.28 \times 10^{7}$                      & $3.37 \times 10^{8}$                      \\
                            &               & Depth of largest circuit        & $2.12 \times 10^{6} $               & $1.35 \times 10^{7}$                      & $1.99 \times 10^{8}$                      \\
\midrule
\midrule
               \textbf{Fault tolerant}             &   Number of qubits            & Largest across circuits        & $35$                      & $35$                      & $35$                      \\
\midrule
                            &     T gates          & Total number across circuits    & $8.62\times10^{6}$                      & $3.49\times10^{8}$                      & $5.85\times10^{9}$                      \\
                            &               & Total depth across circuits     & $7.21\times10^{6}$                      & $2.92\times10^{8}$                      &   $4.89\times10^{9}$                    \\
                            &               & Number in largest circuit       & $3.83\times10^{6}$                    & $9.62\times10^{7}$                      & $1.71\times10^{9}$                      \\
                            &               & Depth of largest circuit        & $3.21\times10^{6}$                      & $8.06\times10^{7}$                      & $1.43\times10^{9}$                      \\
\bottomrule
\end{tabular}
}
\end{table}

\subsection{Separable two-dimensional integration}
\label{sec:separabletwodimensionalintegration}
We now move on to the expression for the actual integration of the tau decay given in \cref{eq:ME}, by also considering the denominator (\emph{i.e.,}\ the propagator). We again omit the integration over the angles. The integral then becomes
\begin{equation}
\label{eq:sep2D}
\sigma \propto \int^{s}_{0} \int^{s-s_2}_{0} \rd s_1 \rd s_2   \frac{s_1^2-s_1 M_{\tau}^2}{(s_2-M^2_\PW)^2 + (M_\PW\Gamma_\PW)^2}.  
\end{equation} The analytical solution to this integral is given in \cref{sec:appendix}. 

In order to implement such a calculation using the QCMI engine, we first note that we can express this integral in the following way
\begin{align}\label{eq:separabletwodimintegration}
\sigma \propto&{} \int^s_0 \int^s_0 \rd s_1 \rd s_2 \frac{1}{(s_2-M_\PW^2)^2 + (M_\PW\Gamma_\PW)^2}s_1^{2} C(s_1,s_2) \nonumber \\
&{}- M_{\tau}^2 \int^s_0 \int^s_0 \rd s_1 \rd s_2 \frac{1}{(s_2-M_W^2)^2 + (M_\PW\Gamma_\PW)^2}s_1 C(s_1,s_2), 
\end{align} 
where the cut function, $C(s_1,s_2)$, is the same as previously. Then, recalling the general form of an expectation calculated using the QMCI engine in \cref{eq:functionappliedintergral2D}, and following the discussion in \cref{sec:generalform}, we see that each individual integral has the canonical form, in terms of products of building blocks, that we require for performing a generic cross-section calculation using the QMCI engine. We can identify the probability distribution, $f_{s_1s_2}$, as a product of univariate BW distributions [or rather, in this case, the product of an uniform distribution for the dimension $s_1$, and an univariate $\PW$-boson BW distribution for the dimension $s_2$, \emph{i.e.}\ $f_{s_1s_2}(s_1,s_2)=U(s_1)\text{BW}_{\PW}(s_2)$]. Then, the functions applied, $g(.)$, are just products of moments of the integration variables [or rather in this case simply the univariate products $g(s_1, s_2)=s_1^{2}$ or $g(s_1,s_2)=s_1$]. 

For this example, for demonstration purposes, it is important that we integrate across the range of the BW peak, and therefore we set $\COM = 100\GeV$.\footnote{Note that this calculation does not correspond to a physical tau decay (or indeed a physical process).} The analytical value in this case is $\sigma = 3.162\times10^{8}$. As discussed in \cref{sec:statepreparation}, we make use of a trained PQC with $n=6$ (specifically the optimal circuit given in \cref{tab:bw_distributions}) to prepare the BW distribution for the $\PW$-boson propagator used to model the probability distributions for dimension $s_2$, and use the QMCI engine's state-preparation library to load a $n=6$ circuit preparing the uniform distribution used to model the probability distribution for dimension $s_1$. This results in a QMCI circuit containing $27$ qubits in total. We again construct $f_{s_1s_2}(s_1,s_2)$ by combining the circuits.

In order to demonstrate the validity of our generic approach, we carried out numerical experiments by running noiseless simulations to numerically estimate the RHS of \cref{eq:separabletwodimintegration}, $24$ different times, for an expected precision of $10\%$. 

\begin{figure}
\centering
\includegraphics[width=0.5\linewidth]{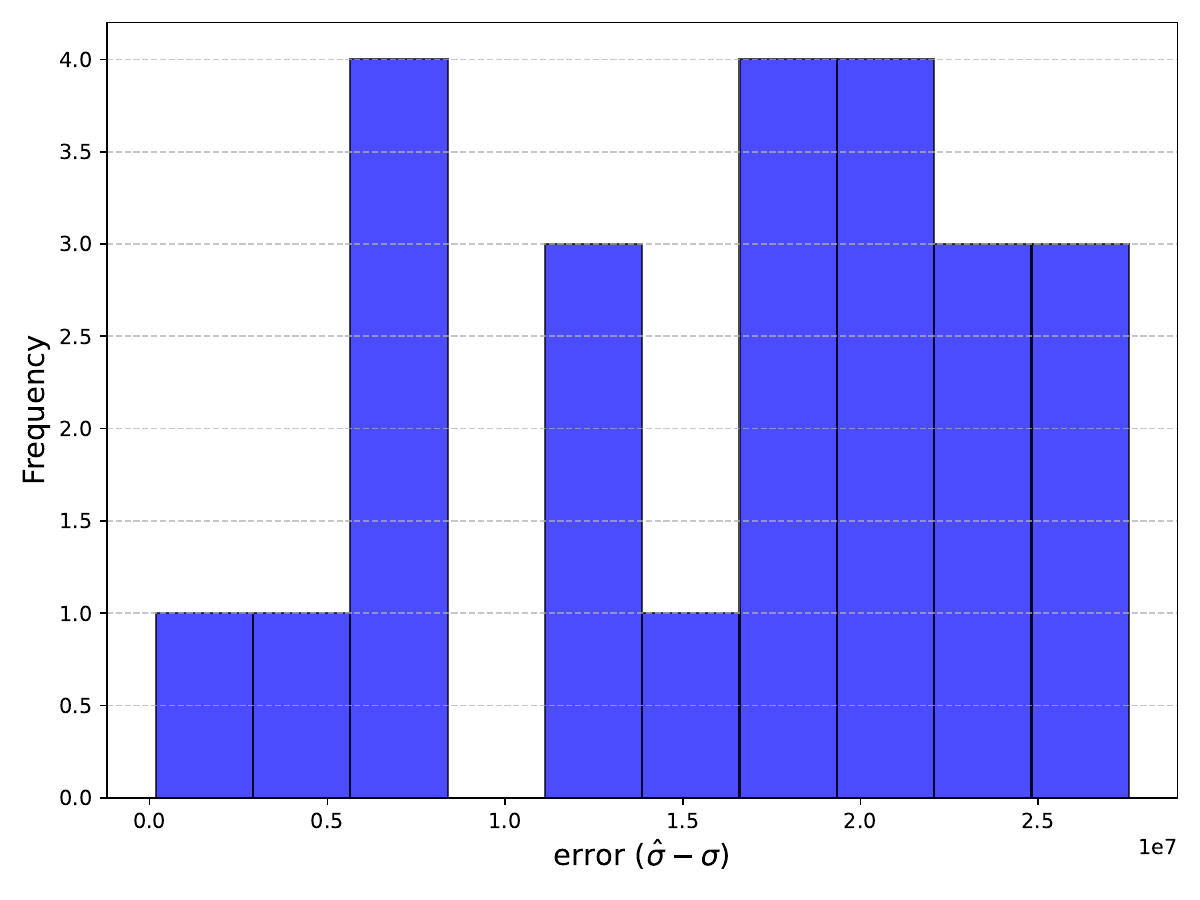}
\caption{Histogram of the error for $24$ runs of QMCI estimating the RHS of \cref{eq:separabletwodimintegration}, for an expected RMSE upper-bound precision of $10\%$.}\label{fig:2dseparableintegrationerror}
\end{figure}

\Cref{fig:2dseparableintegrationerror} gives a histogram of the error, $\hat{\sigma} - \sigma$, where we again see all values within the expected precision of $10\%$. However, in this case there appears to be a bias in the results (as they are not centred around $0$). Some bias is likely expected given the discussion of systematic errors regarding the state preparation of BW distributions given in \cref{sec:statepreparation}---a source of error that was notably not applicable to the previous example. In this case, the observed bias is approximately an order of magnitude smaller than the required precision, and is therefore negligible. However, for higher levels of precision---such as for $1\%$ and $0.1\%$---one would likely require state-preparation circuits with smaller corresponding systematic errors, in order to still obtain accurate results. In particular, a quantitative analysis of the effect of such systematic errors on the final estimation error is needed to be able to fully characterise these QMCI calculations. However, such a detailed analysis would require a significant extension to the error analyses detailed in \cref{sec:statepreparation} (which are derived from Ref.~\cite{cui2024uik}). This is because, for this particular example, systematic errors from the state preparation only indirectly affect the final estimation error via the thresholding condition; the $s_2$ dimension is not directly integrated over, but is summed with the $s_1$ dimension to form a new dimension, and this new dimension used for the threshold condition, thereby indirectly propagating the error in the state preparation to the final estimation. An extension of the error analyses to be able to provide a quantitative treatment for such cases is well beyond the scope of this article, constituting an independent research direction, and is therefore left for future work.

We used the resource mode to calculate the resources for the three expected levels of precision. \cref{tab:resources_2Dseparable} gives the resources for NISQ and fault-tolerant compilation. The results are similar to those found in \cref{sec:onedimensionalintegration}, and therefore a similar analysis to that described there applies. However, in this case it is worth noting that due to the more complicated circuit used to prepare the BW distribution for the $s_2$ dimension, then the qubit counts are slightly larger, and the gate counts are also approximately an order of magnitude larger.

\begin{table}[ht!]
\centering
\caption{Resources required to estimate the RHS of \cref{eq:separabletwodimintegration} using QMCI for various expected precisions, for both NISQ and fault-tolerant resource mode.}
\label{tab:resources_2Dseparable}
\resizebox{\textwidth}{!}{%
\begin{tabular}{lccccc}
\toprule
\textbf{Compilation}           & \textbf{Resource} & \textbf{Metric}                 &  & \textbf{Precision} &  \\
          &  &                 & \textbf{10\%} & \textbf{1\%} & \textbf{0.1\%} \\
\midrule
                        \textbf{NISQ}    &   Number of qubits            & Largest across circuits        & $28$                      & $28$                     & $28$                     \\
\midrule
                    &      CX gates           & Total number across circuits    & $1.34\times10^{7}$             & $1.44\times10^{8}$                      & $1.49\times10^{9}$                    \\
                            &               & Total depth across circuits     & $7.88\times10^{6}$               & $8.43\times10^{7}$                      & $8.74\times10^{8}$                      \\
                            &               & Number in largest circuit       & $4.86\times10^{6}$               &  $6.32\times10^{7}$                     & $7.48\times10^{8}$                     \\
                            &               & Depth of largest circuit        & $2.85\times10^{6}$                 &              $3.71\times10^{7}$          & $4.39\times10^{8}$                      \\
\midrule
                            &    All gates           & Total number across circuits    & $2.72\times10^{7}$               &  $2.91\times10^{8}$                     & $3.02\times10^{9}$                      \\
                            &               & Total depth across circuits     & $1.45\times10^{7}$               &  $1.56\times10^{8}$                     & $1.62\times10^{9}$                      \\
                            &               & Number in largest circuit       & $9.84\times10^{6}$               &  $1.28\times10^{8}$                     & $1.51\times10^{9}$                      \\
                            &               & Depth of largest circuit        & $5.27\times10^{6}$               &      $6.85\times10^{7}$                 & $8.11\times10^{8}$                      \\
\midrule
\midrule
               \textbf{Fault tolerant}             &   Number of qubits            & Largest across circuits        & $41$                      & $41$                      & $41$                      \\
\midrule
                            &     T gates          & Total number across circuits    & $5.37\times10^{8}$                      & $6.97\times10^{9}$                      & $8.23\times10^{10}$                      \\
                            &               & Total depth across circuits     & $5.21\times10^{8}$                      & $6.75\times10^{9}$                      & $7.98\times10^{10}$                      \\
                            &               & Number in largest circuit       & $2.18 \times 10^{8}$                    & $3.08\times10^{9}$                      & $4.21\times10^{10}$                      \\
                            &               & Depth of largest circuit        & $2.11\times10^{8}$                    & $2.99\times10^{9}$                     & $4.08\times10^{10}$                      \\
\bottomrule
\end{tabular}
}
\end{table}

\subsection{Non-separable two-dimensional integration}
\label{sec:fulltwodimensionalintegration}
Finally, in order to increase the complexity of the problem and mimic the more general case of \cref{eq:general_XS}, we compute the following \begin{align}\label{eq:nonseparabletwodimintegration}
\sigma \propto{}& \int^{s}_{0} \int^{s-s_2}_{0} ds_1 ds_2   \frac{s_1^2s-s_1 M_{\tau}^2s + s_1M_{\tau}^2s_2}{(s_2-M^2_\PW)^2 + (M_\PW\Gamma_\PW)^2},
\end{align}
which amounts in practice to, in addition to calculating the same terms as in \cref{eq:separabletwodimintegration} (multiplied by the constant, $s$), calculating the following, non-separable, two-dimensional integral
\begin{align}\label{eq:separabletwodimintegration_I1}
    I_1 ={}& \int^{s}_{0} \int^{s-s_2}_{0} ds_1 ds_2    \frac{s_1 M_\tau^2 s_2}{(s_2-M^2_\PW)^2 + (M_\PW\Gamma_\PW)^2} .
\end{align}
The analytical solution to this additional integral is also provided in \cref{sec:appendix}. The analytical value of \cref{eq:nonseparabletwodimintegration} is then $\sigma = 3.179\times10^{12}$.

For this example, the setup is the same as in \cref{sec:separabletwodimensionalintegration}. In order to demonstrate the validity of our approach, we carried out numerical experiments by running noiseless simulations to numerically estimate the RHS of \cref{eq:nonseparabletwodimintegration}, $6$ times, for an expected precision of $10\%$. 

\begin{figure}
\centering
\includegraphics[width=0.5\linewidth]{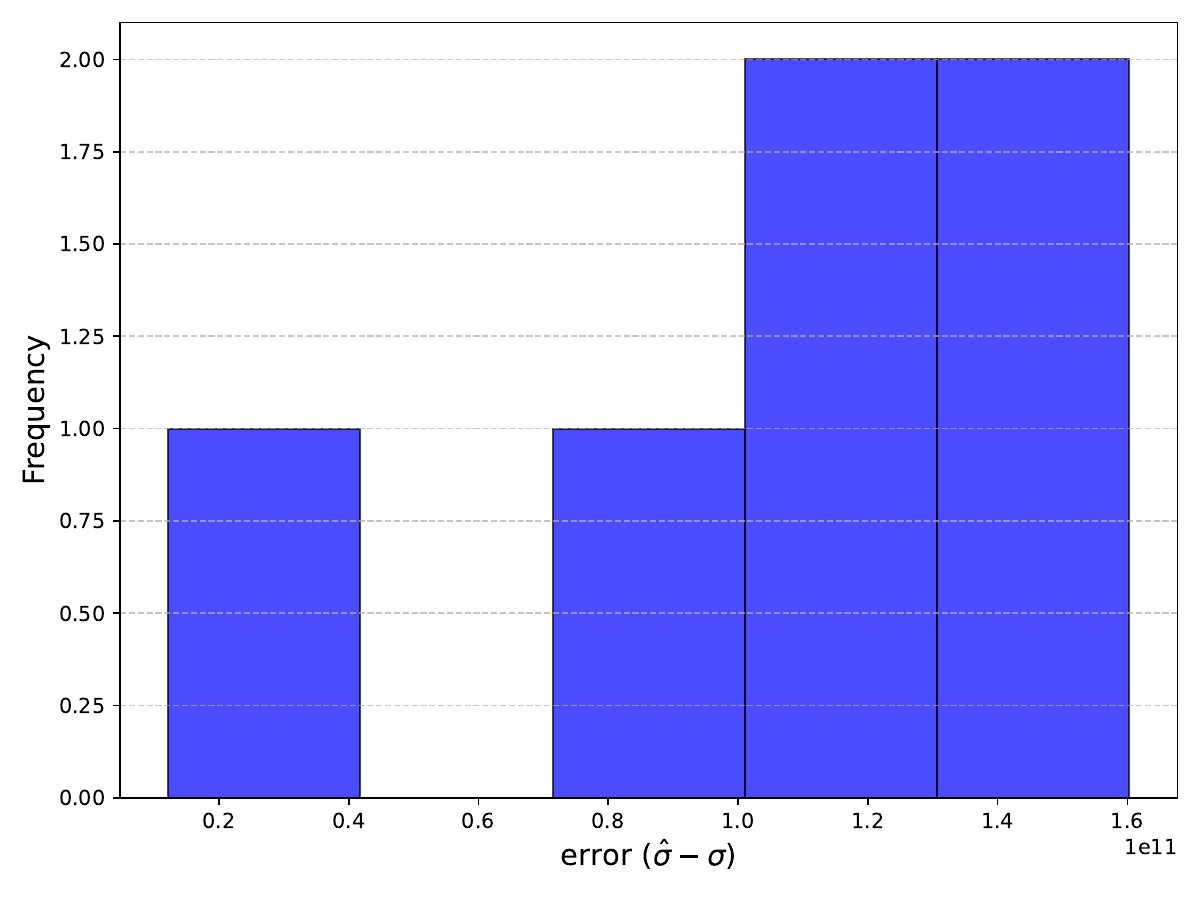}
\caption{Histogram of the error for $6$ runs of QMCI estimating the RHS of \cref{eq:nonseparabletwodimintegration} for an expected RMSE upper-bound precision of $10\%$.}\label{fig:2dnonseparableintegrationerror}
\end{figure}

\Cref{fig:2dnonseparableintegrationerror} gives a histogram of the error, $\hat{\sigma} - \sigma$. The results are broadly similar to those in \cref{sec:separabletwodimensionalintegration}, with all values within the expected precision, and with a bias again observed (this is not surprising, as the calculations only differ by the additional term, which is subdominant).

We analysed the resources required to run the circuits at the nominal precisions using the resource mode. \cref{tab:resources_2D} gives these for NISQ and fault-tolerant compilation. Once again the results are similar to the previous sections, and therefore a similar analysis again applies. In this case, as compared to the example in \cref{sec:separabletwodimensionalintegration}---which itself required greater resources than the example in \cref{sec:onedimensionalintegration}---both the qubit counts and number of gates are larger, by approximately an additional order of magnitude in the latter case. This increase is due to the extra 2D integration that is performed (\emph{i.e.}\ calculating $\mathbb{E}[s_1s_2]$), which as discussed in \cref{sec:fqmci} requires greater resources than when calculating univariate expectations. Thus, we see that as the dimensionality or complexity of the integrals increases, so too do the required resources.

Based upon the resource analysis here, and that in the previous sections, it is clear that to be able to perform state-of-the-art HEP calculations using QMCI in the future---whereby the dimensionality and complexity of the calculations will be much greater than for these examples---we will require significantly greater resources than are available with near-term hardware. Thus, this clearly remains an application for the future, fault-tolerant era of quantum computing.

\begin{table}[ht!]
\centering
\caption{Resources required to estimate the RHS of \cref{eq:nonseparabletwodimintegration} using QMCI for various expected precisions, for both NISQ and fault-tolerant resource mode.}
\label{tab:resources_2D}
\resizebox{\textwidth}{!}{%
\begin{tabular}{lccccc}
\toprule
\textbf{Compilation}           & \textbf{Resource} & \textbf{Metric}                 &  & \textbf{Precision} &  \\
          &  &                 & \textbf{10\%} & \textbf{1\%} & \textbf{0.1\%} \\
\midrule
                        \textbf{NISQ}    &   Number of qubits            & Largest across circuits        & $28$                      & $28$                      & $28$                     \\
\midrule
                    &      CX gates           & Total number across circuits    & $7.39\times10^{7}$            & $6.15\times10^{8}$                      & $5.09\times10^{9}$                    \\
                            &               & Total depth across circuits     & $4.34\times10^{7}$               & $3.61\times10^{8}$                      & $2.99\times10^{9}$                      \\
                            &               & Number in largest circuit       & $3.39\times10^{7}$               & $2.71\times10^{8}$                      & $1.20\times10^{9}$                     \\
                            &               & Depth of largest circuit        & $1.99\times10^{7}$                 & $1.59\times10^{8}$                      & $7.03\times10^{8}$                      \\
\midrule
                            &    All gates           & Total number across circuits    & $1.50\times10^{8}$               & $1.24\times10^{9}$                      & $1.03\times10^{10}$                      \\
                            &               & Total depth across circuits     & $8.02\times10^{7}$               & $6.67\times10^{8}$                      & $5.52\times10^{9}$                      \\
                            &               & Number in largest circuit       & $6.86\times10^{7}$               & $5.49\times10^{8}$                      & $2.42\times10^{9}$                      \\
                            &               & Depth of largest circuit        & $3.68\times10^{7}$               & $2.94\times10^{8}$                      & $1.30\times10^{9}$                      \\
\midrule
\midrule
               \textbf{Fault tolerant}             &   Number of qubits            & Largest across circuits        & $41$                      & $41$                      & $41$                      \\
\midrule
                            &     T gates          & Total number across circuits    & $3.39\times10^{9}$                      & $4.08\times10^{10}$                      & $2.72\times10^{11}$                      \\
                            &               & Total depth across circuits     & $3.29\times10^{9}$                      & $3.95\times10^{10}$                      & $2.63\times10^{11}$                      \\
                            &               & Number in largest circuit       & $1.65\times10^{9}$                    & $2.14\times10^{10}$                      & $6.97\times10^{10}$                      \\
                            &               & Depth of largest circuit        & $1.60\times10^{9}$                     & $2.07\times10^{10}$                     & $6.75\times10^{10}$                      \\
\bottomrule
\end{tabular}
}
\end{table}

\section{Conclusion}
\label{sec:conclusion}

Quantum technology holds the promise of solving complex problems that are classically intractable, or providing more efficient methods for tackling existing challenges than classical approaches allow. Meanwhile, the field of HEP---which seeks to uncover the fundamental laws of nature at the smallest scales---demands immense computational resources. 

In this work, we investigated whether quantum computing could, in the future, help address key computational bottlenecks in HEP. Building on, and going far beyond, the work presented in Ref.~\cite{Agliardi:2022ghn}, where M.P. first introduced the idea of using QMCI for cross-section calculations in HEP, we have made significant advancements in this direction.

We have developed a general approach for integrating non-trivial cross sections in HEP, in terms of decomposing the general integrand into products of constituent building blocks. This work leverages the Fourier QMCI method implemented in {\sc Quantinuum}’s QMCI engine~\cite{Akhalwaya:2023hqe} (developed in part by I.W.) along with several other key proprietary features of the engine. Specifically, we introduced two distinct approaches for preparing relativistic BW distributions on quantum registers---functions that appear as one of the key building blocks of generic integrands in cross-section calculations. 

To demonstrate the method’s capabilities, we performed two-dimensional integrations for several examples that arguably represent some of the most challenging integrands investigated to date for quantum integration. However, it is still worth remarking that these examples remain significantly less complex than the high-dimensional integrands encountered in state-of-the-art classical calculations.

Our findings suggest that such applications are unlikely to be practical during the NISQ era. Instead, they are expected to become viable when fault-tolerant quantum devices are available. Notably, the resource estimates we provide for fault tolerance represent loose upper bounds. We anticipate that significant improvements will be achieved in the future---and thus significantly reduced resource requirements---as the field of fault-tolerant compilation matures beyond its infancy.

While this work represents a significant first step toward the quantum integration of cross sections in HEP, there are several areas for improvement. One limitation is the use of uniform spacing for representing underlying probability distributions with qubits, which is an inherent requirement of the QMCI engine. Extending the engine’s functionality to allow for non-uniform spacing could significantly reduce both the number of qubits and the quantum gates required for integration. Additionally, while the method is general and theoretically applicable to integrals of arbitrary dimensionality (beyond the two-dimensional examples studied here), its efficiency will decrease for high-dimensional problems due to the polynomial scaling in the number of terms to calculate. Adapting the method to achieve more favourable scaling would enhance its practicality, and studying how circuit size scales heuristically with integration dimensionality would be particularly valuable.

Another avenue for improvement concerns the handling of experimental constraints, as discussed in \cref{sec:XsectionHEP}, as these could be considered in further generality. Restricted integration domains based on constraints such as sums, maximum/minimum etc., can be easily implemented via the QMCI engine’s enhanced P-builder functionality. However, in more general cases, the cut functions may be much more complex, or even non-analytical, and this thus requires further exploration to ensure efficient integration for these cases. Similarly, extending the method to accommodate more general forms of cross-section integrands would broaden its applicability. Accurate preparation of relativistic BW distributions, which is crucial to this method, represents another area for further investigation. Alternative state-preparation methods that are tailored to the particular distribution could enhance both accuracy and resource efficiency, and also could be of use in general to the HEP community for other future quantum computing applications where resonances are required to be modelled. In particular, understanding in detail the systematic errors associated with state preparation on the final estimation error is critical for performing precise integrations, and this issue warrants more detailed exploration.
Finally, event sampling based on underlying distributions is a ubiquitous task in HEP. Exploring whether quantum computing and quantum integration could enhance efficiency or accuracy in this area would be of particular interest, and could provide significant advancements for the field.
In summary, while this work lays a solid foundation for quantum integration in HEP, addressing these challenges will significantly expand its potential and applicability.

\section*{Acknowledgements}
The authors would like to thank Matthew DeCross and Julien Sorci for their careful reviewing of the manuscript.
The authors thank Steven Herbert and Alexandre Krajenbrink for their support in this work and interesting discussions (and the latter for also reviewing the manuscript).
The authors also want to thank Matthias Rosenkranz for interesting discussions.
M.P. acknowledges support by the German Research Foundation (DFG) through the Research Training Group RTG2044.

\appendix
\section{Appendix}
\label{sec:appendix}

\subsection{State preparation of relativistic BW distributions}\label{app:stateprep}

In this section, we have some additional plots related to state preparation of BW distributions.

\begin{figure}[ht!]
    \centering
    \begin{subfigure}{0.49\textwidth}
        \centering \includegraphics[width=0.95\linewidth]{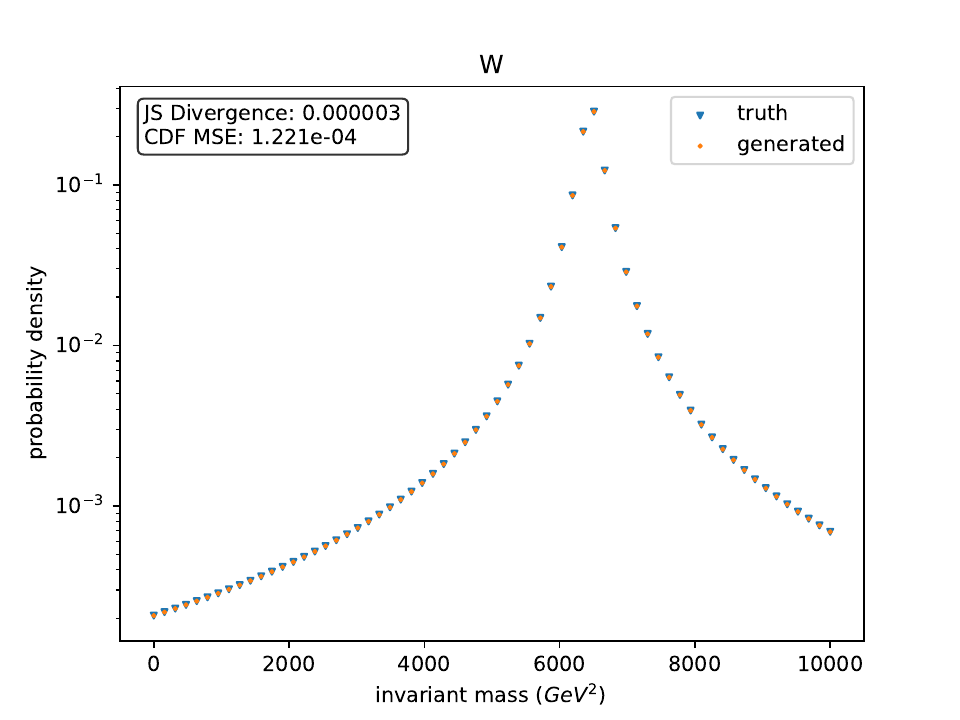}
        (a)
    \end{subfigure}
    \begin{subfigure}{0.49\textwidth}
        \centering \includegraphics[width=0.95\linewidth]{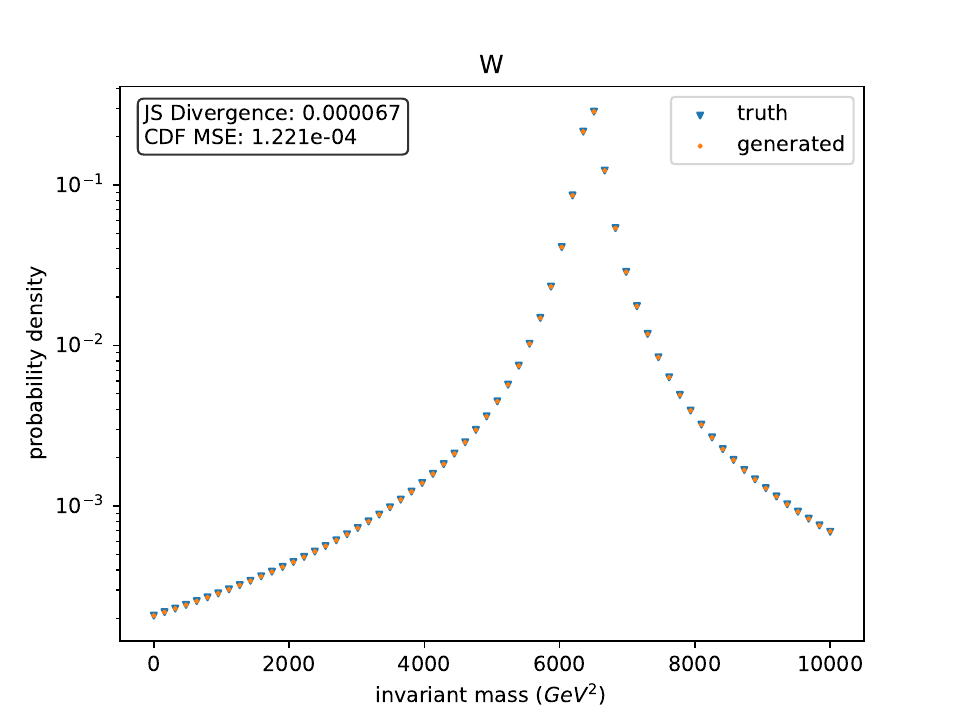}
        (b) $d=200$
    \end{subfigure}
    \caption{True (blue triangles) and generated (orange circles) points for the $\PW$-boson BW distribution up to $\COM=100 \GeV$, generated using (left) the variational method and (right) the Fourier expansion method.}
    \label{fig:w_100GeV}
    \end{figure}

\subsection{Analytical expressions}

In this section, we provide the analytical expressions for several integrals required in the main text.

First, the analytical expression for the integral in \cref{eq:sep2D} in \cref{sec:separabletwodimensionalintegration} reads
\begin{align}
I = {}&\int^{s}_{0} \int^{s-s_2}_{0}  \rd s_1 \rd s_2 \, \frac{s_1^2-s_1 M_\tau^2}{(s_2-M^2_\PW)^2 + (M_\PW\Gamma_\PW)^2}  \nonumber\\
= {}& \frac{1}{6 \Gamma_{\PW} M_{\PW}} \left[\Gamma_{\PW} M_\PW \left(\Gamma_{\PW}^2 M_\PW^2-3
   \left(M_\PW^2-s\right) \left(M_\tau^2+M_\PW^2-s\right)\right)
   \log \left[\frac{\Gamma_{\PW}^2
   M_\PW^2+\left(M_\PW^2-s\right)^2}{M_\PW^2
   \left(\Gamma_{\PW}^2+M_\PW^2\right)}\right] \right. \nonumber \\
   &{} \left. +\tan
   ^{-1}\left(\frac{M_\PW}{\Gamma_{\PW}}\right) \left(3 \Gamma_{\PW}^2
   M_\PW^2 \left(M_\tau^2+2 M_\PW^2-2
   s\right)-\left(M_\PW^2-s\right)^2 \left(3 M_\tau^2+2 M_\PW^2-2
   s\right)\right) \right. \nonumber \\
   &{} \left. +\left(\left(M_\PW^2-s\right)^2 \left(3 M_\tau^2+2
   M_\PW^2-2 s\right)-3 \Gamma_{\PW}^2 M_\PW^2 \left(M_\tau^2+2
   M_\PW^2-2 s\right)\right) \cot ^{-1}\left(\frac{\Gamma_{\PW}
   M_\PW}{M_\PW^2-s}\right) \right. \nonumber \\
   &{} \left. +\Gamma_{\PW} M_\PW s \left(-3
   M_\tau^2-4 M_\PW^2+5 s\right) \right] .
\end{align}
Then, the analytical result for the additional integral in \cref{eq:separabletwodimintegration_I1} in \cref{sec:fulltwodimensionalintegration} reads
%
\begin{align}
    I_1 ={}& \int^{s}_{0} \int^{s-s_2}_{0} \rd s_1 \rd s_2 \, \frac{s_1 M_\tau^2 s_2}{(s_2-M^2_\PW)^2 + (M_\PW\Gamma_\PW)^2} \nonumber\\
    ={}& \frac{M_\tau^2 }{4 \Gamma_\PW} \bigg[-2 M_\PW \left(\Gamma_\PW^2 \left(2 s-3
   M_\PW^2\right)+\left(M_\PW^2-s\right)^2\right) \left(\cot
   ^{-1}\left(\frac{\Gamma_\PW M_\PW}{M_\PW^2-s}\right)-\tan
   ^{-1}\left(\frac{M_\PW}{\Gamma_\PW}\right)\right) \nonumber \\
   {}& +\Gamma_\PW
   \left(\Gamma_\PW^2 M_\PW^2-3 M_\PW^4+4 M_\PW^2 s-s^2\right) \log
   \left(\frac{M_\PW^2 \left(\Gamma_\PW^2+M_\PW^2\right)}{\Gamma_\PW^2
   M_\PW^2+\left(M_\PW^2-s\right)^2}\right)+\Gamma_\PW s \left(4
   M_\PW^2-3 s\right)\bigg] . 
\end{align}

\bibliographystyle{utphys.bst}
\providecommand{\href}[2]{#2}\begingroup\raggedright\endgroup

\end{document}